\documentclass[12pt,letterpaper]{article}
\pdfoutput=1
\usepackage{amssymb}
\usepackage{amsmath}
\usepackage{fourier-orns}
\usepackage{amscd}
\usepackage{graphicx}
\usepackage{hyperref}
\def\be{\begin{equation}}
\def\ee{\end{equation}}
\def\frc#1#2{\relax\ifmmode{\textstyle\frac{#1}{#2}} 
                    \else$\frac{#1}{#2}$\fi}         

\def\rI{{{}_{\rm I}}}
\def\rJ{{{}_{\rm J}}}

\def\hj{{\hat\jmath}}
\def\hk{{\hat k}}
\def\hi{{\hat\imath}}

\def\fracm#1#2{\hbox{\large{${\frac{{#1}}{{#2}}}$}}}

\def\vCent#1{\vcenter{\hbox{\hss#1\hss}}}

\def\hi{{\hat\imath}}
\def\hj{{\hat\jmath}}
\def\hk{{\hat{k}}}
\def\hl{{\hat\ell}}

\def\pp{{\mathchoice
          {
              \kern 1pt%
              \raise 1pt
              \vbox{\hrule width5pt height0.4pt depth0pt
                    \kern -2pt
                    \hbox{\kern 2.3pt
                          \vrule width0.4pt height6pt depth0pt
                          }
                    \kern -2pt
                    \hrule width5pt height0.4pt depth0pt}%
                    \kern 1pt
           }
            {
              \kern 1pt%
              \raise 1pt
              \vbox{\hrule width4.3pt height0.4pt depth0pt
                    \kern -1.8pt
                    \hbox{\kern 1.95pt
                          \vrule width0.4pt height5.4pt depth0pt
                          }
                    \kern -1.8pt
                    \hrule width4.3pt height0.4pt depth0pt}%
                    \kern 1pt
            }
            {
              \kern 0.5pt%
              \raise 1pt
              \vbox{\hrule width4.0pt height0.3pt depth0pt
                    \kern -1.9pt  
                    \hbox{\kern 1.85pt
                          \vrule width0.3pt height5.7pt depth0pt
                          }
                    \kern -1.9pt
                    \hrule width4.0pt height0.3pt depth0pt}%
                    \kern 0.5pt
            }
            {
              \kern 0.5pt%
              \raise 1pt
              \vbox{\hrule width3.6pt height0.3pt depth0pt
                    \kern -1.5pt
                    \hbox{\kern 1.65pt
                          \vrule width0.3pt height4.5pt depth0pt
                          }
                    \kern -1.5pt
                    \hrule width3.6pt height0.3pt depth0pt}%
                    \kern 0.5pt
            }
        }}

\def\mm{{\mathchoice
                       {
                             \kern 1pt
               \raise 1pt    \vbox{\hrule width5pt height0.4pt depth0pt
                                  \kern 2pt
                                  \hrule width5pt height0.4pt depth0pt}
                             \kern 1pt}
                       {
                            \kern 1pt
               \raise 1pt \vbox{\hrule width4.3pt height0.4pt depth0pt
                                  \kern 1.8pt
                                  \hrule width4.3pt height0.4pt depth0pt}
                             \kern 1pt}
                       {
                            \kern 0.5pt
               \raise 1pt
                            \vbox{\hrule width4.0pt height0.3pt depth0pt
                                  \kern 1.9pt
                                  \hrule width4.0pt height0.3pt depth0pt}
                            \kern 1pt}
                       {
                           \kern 0.5pt
             \raise 1pt  \vbox{\hrule width3.6pt height0.3pt depth0pt
                                  \kern 1.5pt
                                  \hrule width3.6pt height0.3pt depth0pt}
                           \kern 0.5pt}
                       }}

\def\ad{{\kern0.5pt
                   \alpha \kern-5.05pt \raise5.8pt\hbox{$\textstyle.$}\kern
0.5pt}}

\def\bd{{\kern0.5pt
                   \beta \kern-5.05pt \raise5.8pt\hbox{$\textstyle.$}\kern
0.5pt}}

\def\qd{{\kern0.5pt
                   q \kern-5.05pt \raise5.8pt\hbox{$\textstyle.$}\kern
0.5pt}}
\def\Dot#1{{\kern0.5pt
     {#1} \kern-5.05pt \raise5.8pt\hbox{$\textstyle.$}\kern
0.5pt}}

\catcode`@=11
\def\un#1{\relax\ifmmode\@@underline#1\else
        $\@@underline{\hbox{#1}}$\relax\fi}
\catcode`@=12

\def\a{\alpha}
\def\b{\beta}

\def\g{\gamma}

\def\k{\kappa}
\def\l{\lambda}
\def\m{\mu}

\def\t{\tau}

\def\D{\Delta}

\def\L{\Lambda}

\def\dslash{\not{\hbox{\kern-2pt $\partial$}}}
\def\Dslash{\not{\hbox{\kern-4pt $D$}}}
\def\pslash{\not{\hbox{\kern-2.3pt $p$}}}
 \newtoks\slashfraction
 \slashfraction={.13}
 \def\slash#1{\setbox0\hbox{$ #1 $}
 \setbox0\hbox to \the\slashfraction\wd0{\hss \box0}/\box0 }
 
\font\ro=cmsy10                          
\def\kcr{{\hbox{\ro \char'170}}}                
\def\ktl{{\hbox{\ro \char'170}}}        
\def\ktr{{\hbox{\ro \char'170}}}        
\def\kbl{{\hbox{\ro \char'170}}}        
\def\kbr{{\hbox{\ro \char'170}}}        

\def\plpl{\raise-2pt\hbox{$\raise3pt\hbox{$_+$}\hskip-6.67pt\raise0.0pt
\hbox{$^+$}\hskip 0.01pt$}}
\def\mimi{\raise-2pt\hbox{$\raise3pt\hbox{$_-$}\hskip-6.67pt\raise0.0pt
\hbox{$^-$}\hskip 0.01pt$}} 

\def\bo{{\raise.15ex\hbox{\large$\Box$}}}               
\def\pa{\partial}                                       
\def\iff{\leftrightarrow}                               
\def\TH{{\raise.2ex\hbox{$\displaystyle \bigodot$}\mskip-4.7mu \llap H \;}}
\def\face{{\raise.2ex\hbox{$\displaystyle \bigodot$}\mskip-2.2mu \llap {$\ddot
        \smile$}}}                                      

\def\Hat#1{\widehat{#1}}                        
\def\leftrightarrowfill{$\mathsurround=0pt \mathord\leftarrow \mkern-6mu
        \cleaders\hbox{$\mkern-2mu \mathord- \mkern-2mu$}\hfill
        \mkern-6mu \mathord\rightarrow$}
\def\dvec#1{\vbox{\ialign{##\crcr
        \leftrightarrowfill\crcr\noalign{\kern-1pt\nointerlineskip}
        $\hfil\displaystyle{#1}\hfil$\crcr}}}           
\def\dt#1{{\buildrel {\hbox{\LARGE .}} \over {#1}}}     

\def\fracm#1#2{\hbox{\large{${\frac{{#1}}{{#2}}}$}}}
\def\frac#1#2{{\textstyle{#1\over\vphantom2\smash{\raise.20ex
        \hbox{$\scriptstyle{#2}$}}}}}                   
\def\sfrac#1#2{{\vphantom1\smash{\lower.5ex\hbox{\small$#1$}}\over
        \vphantom1\smash{\raise.4ex\hbox{\small$#2$}}}} 
\def\bfrac#1#2{{\vphantom1\smash{\lower.5ex\hbox{$#1$}}\over
        \vphantom1\smash{\raise.3ex\hbox{$#2$}}}}       
\def\afrac#1#2{{\vphantom1\smash{\lower.5ex\hbox{$#1$}}\over#2}}    
\def\on#1#2{\mathop{\null#2}\limits^{#1}}               

\def\opR{{\bm{\cal R}}}

\def\pa{\partial}      
\newcommand{\bm}[1]{\mbox{\boldmath$#1$}}

\def\ad{{\dot{\alpha}}}
\def\bd{{\dot{\beta}}}

\font\ro=cmsy10                          
\def\kcr{{\hbox{\ro \char'170}}}                
\def\ktl{{\hbox{\ro \char'170}}}        
\def\ktr{{\hbox{\ro \char'170}}}        
\def\kbl{{\hbox{\ro \char'170}}}        
\def\kbr{{\hbox{\ro \char'170}}}        

\parskip=\medskipamount                 
\thicklines                         

\def\border{                                            
        \setlength{\unitlength}{1mm}
        \newcount\xco
        \newcount\yco
        \xco=-21
        \yco=12
        \begin{picture}(140,0)
        \put(\xco,\yco){$\ktl$}
        \advance\yco by-1
        {\loop
        \put(\xco,\yco){$\kcr$}
        \advance\yco by-2
        \ifnum\yco>-240
        \repeat
        \put(\xco,\yco){$\kbl$}}
        \xco=158
        \yco=12
        \put(\xco,\yco){$\ktr$}
        \advance\yco by-1
        {\loop
        \put(\xco,\yco){$\kcr$}
        \advance\yco by-2
        \ifnum\yco>-240
        \repeat
        \put(\xco,\yco){$\kbr$}}
        \put(-20,13){\tiny **University of Maryland * Center for String and
         Particle  Theory* Physics Department***University of Maryland *Center
        for String and Particle  Theory** }
        \put(-20,-241.5){\tiny **University of Maryland * Center for String and
         Particle  Theory* Physics Department***University of Maryland *Center
        for String and Particle  Theory** }
        \end{picture}
        \par\vskip-8mm}

\def\headpic{                                           
        \indent
        \setlength{\unitlength}{.4mm}
        \thinlines
        \par
        \begin{picture}(29,16)
        \put(165,16){\line(1,0){4}}
        \put(170,16){\line(1,0){4}}
        \put(180,16){\line(1,0){4}}
        \put(175,0){\line(1,0){4}}
        \put(180,0){\line(1,0){4}}
        \put(185,0){\line(1,0){4}}
        \put(169,0){\line(0,1){16}}
        \put(170,0){\line(0,1){16}}
        \put(179,0){\line(0,1){16}}
        \put(180,0){\line(0,1){16}}
        \put(184,0){\line(0,1){16}}
        \put(185,0){\line(0,1){16}}
        \put(169,16){\oval(8,32)[bl]}
        \put(170,16){\oval(8,32)[br]}
        \put(179,0){\oval(8,32)[tl]}
        \put(185,0){\oval(8,32)[tr]}
        \end{picture}
        \par\vskip-6.5mm
        \thicklines}
\def\endtitle{\end{quotation}\newpage}                  

\newskip\humongous \humongous=0pt plus 1000pt minus 1000pt
\def\caja{\mathsurround=0pt}
\def\eqalign#1{\,\vcenter{\openup2\jot \caja
        \ialign{\strut \hfil$\displaystyle{##}$&$
        \displaystyle{{}##}$\hfil\crcr#1\crcr}}\,}
\newif\ifdtup

\renewcommand{\=}{~=~}

\renewcommand{\L}{{\rm L}}
\newcommand{\R}{{\rm R}}
\renewcommand{\D}{{\rm D}}

\begin{document}
\numberwithin{equation}{section} 
\setcounter{page}{0}
\thispagestyle{empty}

\def\dt#1{\on{\hbox{\bf .}}{#1}}                
\def\Dot#1{\dt{#1}}

\def\gfrac#1#2{\frac {\scriptstyle{#1}}
        {\mbox{\raisebox{-.6ex}{$\scriptstyle{#2}$}}}}
\def\gg{{\hbox{\sc g}}}
\border\headpic {\hbox to\hsize{\today \hfill
{UMDEPP-014-004}}}
\par \noindent
\par

\setlength{\oddsidemargin}{0.3in}
\setlength{\evensidemargin}{-0.3in}
\begin{center}
\vglue .08in
{\large\bf Is It Possible To Embed A \\ 
\vskip.1in 4D, $\bm {\cal N}$
= 4 Supersymmetric Vector Multiplet
\vskip.1in Within A Completely Off-Shell Adinkra Hologram?}\\[.8in]

Mathew Calkins\footnote{mathewpcalkins@gmail.com}, D.\ E.\ A.\ 
Gates\footnote{deagates@terpmail.umd.edu}, S.\, 
James Gates, Jr.\footnote{gatess@wam.umd.edu}${}^{\dagger}$,
and Brian McPeak\footnote{bmcpeak@terpmail.umd.edu}
\\[0.3in]
${}^\dag${\it Center for String and Particle Theory\\
Department of Physics, University of Maryland\\
College Park, MD 20742-4111 USA}\\[1in]
 $~$
\\[.3in]
{\bf ABSTRACT}\\[.01in]
\end{center}
\begin{quotation}
{We present evidence of the existence of a 1D, $N$ = 16 SUSY
hologram that can be used to understand representation theory 
aspects of a 4D, $\cal N$ = 4 supersymmetrical vector multiplet. 
In this context, the long-standing ``off-shell SUSY'' problem for the 
4D, $\cal N$ = 4 Maxwell supermultiplet is precisely formulated 
as a problem in linear algebra.}
\\[.3in]
\vfill
\endtitle

\setlength{\oddsidemargin}{0.3in}
\setlength{\evensidemargin}{-0.3in}

\setcounter{equation}{0}
\section{Introduction}

$~~~$ The utility of the extra dimension concept underwent a 
reassessment due to the construction of 11D
supergravity \cite{SG11D1,SG11D2}.  Since then, the 
concept has received much attention even to the point of producing a broadly
studied approach \cite{RS1,RS2,RS3} (e.\ g.\ `brane-world scenarios') 
that dominated phenomenological discussion for a decade.  The 11D 
approach was the solution to a difficult problem
as Cremmer 
and Julia used it to present the first complete description of 4D, 
$\cal N$ = 8 supergravity \cite{SG4dn8}.

Thus, for perhaps the first time in the literature associated with supergravity,
it was shown that a higher dimensional approach contained information
about a lower dimensional theory that could be more easily accessed from 
the higher dimensional starting point.  This same idea can be seen in the
relation of 10D, $\cal N$ = 1 SUSY YM theories to 4D, $\cal N$ = 4 SUSY 
YM theories.  The key point to note is that the information necessary for 
the construction of both the higher dimensional and lower dimensional 
theories is conserved by either the dimensional reduction or dimensional 
extension processes.  The lower dimensional theory acts as a hologram for 
the higher dimensional one.  These facts are well known.

Starting in 1994 \cite{GRana0}, we began to find evidence \cite{GRana1,
GRana2,GRana3,GRana4}
that this well known result extends all the way from supersymmetric 
quantum field theories to supersymmetric quantum mechanical models 
and more unexpectedly the conservation of the information may be so robust 
that it might allow the former to be re-constructed from the latter in some
limits.  Eventually
we gave this idea a name ``SUSY holography'' \cite{ENUF1,ENUF2}.

The topic of the 4D, $\cal N$ = 4 Yang-Mills supermultiplet \cite{N4YMo1,
N4YMo2,N4YMo3} 
has been a fruitful one for many years.  Almost from the instant of its first 
presentation, the unusual properties of this model have generated a steady 
stream of inspirations concluding most recently with the introduction of the 
``amplituhedron''  \cite{amplituH1,amplituH2}.  Thus, this theory has long been one of 
our objectives to study via the tools that have been developed for SUSY 
holography.

Stated another way, it is our goal to follow a path similar to that of Cremmer
and Julia, but to use the idea of ``SUSY holography'' in the reverse route of 
using a lower dimensional construct, at least at the level of representation 
theory, to gain greater understanding of a higher dimensional construct.  One 
of our previous works \cite{N4YM1}, presented (what may be the most) 
detailed results on the nature of the non-closure of the SUSY algebra for 
the 4D, $\cal N$ = 4 supermultiplet in the context of an equivalent $\cal N$ 
= 1 superfield formulation solely in four dimensions.  In this current work, we 
will begin the process of studying its projection (or shadow) into the sea of 
one dimensional $N$ = 16 adinkra networks known to exist.

\section{The SUSY Holography Conjecture}

$~~~$ On first reflection, the proposal of ``SUSY holography'' would seem 
untenable.  There is an easy example to show why this conclusion might 
be reached.  For a four dimensional field theory (supersymmetrical or not), 
the starting point for a reduction to one dimension can be implemented 
by making the replacement
\be  \eqalign{
\pa_{\m}
 ~=~ {\cal T}_{\mu} \, {{\pa} \over {\partial \t}} 
~~~,~~~
{\cal T}_{\mu}  ~\equiv~ (\, 1, \,0,\, 0 ,\, 0 \, ) ~~~,
}    \label{0brane}
\ee
in actions.  As well, all field variables are assumed to depend only on 
the real parameter $\t$ and all gauge fields are restricted to the Coulomb 
gauge.  In the context of a non-supersymmetrical theory, such a reduction 
can lead to an ambiguity involving a loss of spin-bundle information.  Let 
us consider two distinct four dimensional theories: \newline (a.) an action 
involving three parity-even massless spin-0 fields $\phi^{\cal I}$, (where 
$\cal I$ = 1, 2,
\newline $~~~~~~$ 
and 3)
\be\eqalign{
{\mathcal{L}}_{Spin-0}  ~=~ &-\frac{1}{2}(\partial_{\mu}\phi^{\cal I})(\partial^{\mu}\phi^{\cal I})  ~~~,
}  \label{act1} \ee $~~~~~~$ 
so that under the prescription of (\ref{0brane}) leads to
\be\eqalign{
{\mathcal{L}}_{Spin-0}  ~=~ &\frac{1}{2}(\partial_{\t}\phi^{1})(\partial_{\t}\phi^{1})
\,+\, \frac{1}{2}(\partial_{\t}\phi^{2})(\partial_{\t}\phi^{2})  \,+\,  
\frac{1}{2}(\partial_{\t}\phi^{3})(\partial_{\t}\phi^{3})
 ~~~, ~~{\rm {and}}
}  \label{act1b} \ee
(b.) an action for a spin-1 gauge field given by
\be\eqalign{
{\mathcal{L}}_{Spin-1} ~=~ 
&-\frac{1}{4}\, F_{\mu\nu}F^{\mu\nu} ~=~ -\frac{1}{2}\, F_{0 \, i}F^{0 \, i}  ~~~.
}  \label{act2} \ee $~~~~~~$ 
(since all spatial derivatives vanish in our reduction scheme) and following the 
\newline $~~~~~~~$ 
prescription above this becomes,
\be\eqalign{
{\mathcal{L}}_{Spin-1}  
~=~ &\frac{1}{2} \, \left[ \, (\,  \partial_\tau  \, A{}_{1} \, )^2 ~+~ 
(\,  \partial_\tau  \, A{}_{2} \, )^2 ~+~ (\,  \partial_\tau  \, A{}_{3} \, )^2 
\, \right] ~~~.
}  \label{act2b} \ee

As is seen above, the forms of (\ref{act1b}) and (\ref{act2b}) are exactly the 
same.  Thus, starting from a non-supersymmetrical one dimensional theory
involving three bosonic fields there is no way to distinguish which of the four 
dimensional actions were its origin.  Specifically, the information on the 4D 
spin-bundle of the fields was lost.  This is an example of what we refer to 
as loss of information under non-supersymmetric 0-brane reduction described 
by (\ref{0brane}).

Remarkably, within the context of supersymmetrical theories, this information
can be conserved...if one looks into the ``correct'' structure.

While it is true that the information about the 4D origins of the actions does
not appear in either 1D action, if these non-supersymmetrical theories are
embedded within 1D, $N$ = 4 theories at one extreme and 4D, $\cal N$ = 1 
theories on the other, the information can be subtly encoded in the SUSY 
variations!

We begin with the
spin-0 field and for the sake of simplicity, we only need to consider a single
such field.  Since it has parity-even, it becomes the $A$-field in a chiral
supermultiplet as part of the collection of fields $\left(  A, \, B, \, \psi_a , \, F, \, G  \right)$.
In a similar manner, the spatial vector $\vec A$ can be combined
with its temporal component $A_0$ to form a 4-vector $A_{\m}$ and 
becomes the gauge field in a vector supermultiplet among the collection
of fields $\left( \, A_{\m}, \, \lambda_a , \, {\rm d} \, \right)$. So these are
the relevant 4D supermultiplets.

In 1D, a valise formulation that is off-shell is one where a set of bosonic variables $\Phi_i(\tau)$
and fermionic variables $ \Psi_{\hat{k}} (\tau)$ {\em {locally}} satisfy 
the following realization under the action of a set of supercovariant 
derivatives $\D_{\rm I}$
\be
\D_{\rm I} \Phi_i \= i\, (\L_{\rm I})_{i \hat{k}} \Psi_{\hat{k}} \qquad \text{and} 
\qquad \D_{\rm I} \Psi_{\hat{k}} \= (\R_{\rm I})_{\hat{k} i} \, \left( \pa_{ \tau}
 \Phi_i  \right)
~~~,
 \label{CM9}
\ee
with L-matrices and R-matrices satisfying 
\be \eqalign{
 (\,{\rm L}_\rI\,)_i{}^\hj\>(\,{\rm R}_\rJ\,)_\hj{}^k + (\,{\rm L}_\rJ\,)_i{}^\hj\>(\,{\rm 
 R}_\rI\,)_\hj{}^k &= 2\,\delta_{\rI\rJ}\,\delta_i{}^k~~,\cr
 (\,{\rm R}_\rJ\,)_\hi{}^j\>(\, {\rm L}_\rI\,)_j{}^\hk + (\,{\rm R}_\rI\,)_\hi{}^j\>(\,{\rm 
 L}_\rJ\,)_j{}^\hk
  &= 2\,\delta_{\rI\rJ}\,\delta_\hi{}^\hk~~.
}  \label{GarDNAlg1}
 \ee
\be
~~~~
 (\,{\rm R}_\rI\,)_\hj{}^k\,\delta_{ik} = (\,{\rm L}_\rI\,)_i{}^\hk\,\delta_{\hj\hk}~~,
\label{GarDNAlg2}
\end{equation}
and where the indices range as $\rm I$, $\rm J$, etc.  = 1, $\dots$, $
N$; i, j, etc. = 1, $\dots$, d ; and $\hi$, $\hj$, etc. =  1, 
$\dots$, d for integers d, and $N$.

Implementation of the reduction process described at the beginning of this
section is not sufficient to arrive at a valise formulation of these 4D, $\cal N$
= 1 supermultiplets.  In order to obtain a valise will also require that we make 
the `field redefinitions' 
\be
F^{\cal I} ~\to~ \pa_{\tau} F^{\cal I}  ~~,~~ G^{\cal I} ~\to~ \pa_{\tau} G^{\cal I}   
 ~~,~~ {\rm d}~\to~ \pa_{\tau} {\rm d} ~~,
\label{ReDef}
\ee
after the reduction.  In a subsequent chapter, we will obtain the valise formulation
of the 4D, $\cal N$ = 4 theory as the main new result of this work.

Applying all of this machinery to the components of a chiral supermultiplet
we find
\be \eqalign{
{\rm D}_a A  & \= \psi_a  ~~~~~~~~~~\,~~~~,~~~
{\rm D}_a B   \= i\, ( \gamma^5 )_a{}^b \psi_b  ~~~~~~~~~~,~~~  \cr
{\rm D}_a F  & \= ( \gamma \cdot {\cal T})_a{}^b \, \psi_b  ~~~, ~~~
{\rm D}_a G   \= i\, ( \gamma^5 \gamma \cdot {\cal T} )_a{}^b \, \psi_b   
~~~, \cr
{\rm D}_a \psi_b  & \= i\, (\gamma \cdot {\cal T} )_{ab}  \left( \,\partial_{
\tau} A  \, \right) - ( \gamma^5 \gamma \cdot {\cal T})_{ab}  \left( \, 
\partial_{\tau} B  \, \right) - i C_{ab}  \left( \,  \partial_{\tau} F  \, \right) + 
( \gamma^5 )_{ab}  \left( \, \partial_{\tau} G   \, \right) \, ~,
}
\label{CMv}
\ee
and in a similar manner for the components of the vector supermultiplet,
one is led to
\be \eqalign{
\D_a A_i & ~=~ (\gamma_i)_a{}^b \lambda_b  ~~~, ~~~ D_a {\rm d} 
 \, =\,  i (\gamma^5 \gamma \cdot {\cal T})_a{}^b \, \lambda_b 
 ~~~~~~~~~~~~~~, \cr
\D_a \lambda_b & ~=~ - \tfrac{i}{2}( [ \, \gamma \cdot {\cal T} ~,~ 
\gamma^i \, ])_{ab} \,  \left( \,  \pa_{\tau} A_i  \, \right)~+~ (\gamma^5
)_{ab} \,  \left( \, \pa_{\tau} {\rm d} \, \right)
~~~.  }    \label{VMv}
\ee
All the equations in (\ref{CMv}) and (\ref{VMv}) have exactly the
form of (\ref{CM9}).

Under the reduction described above, there is a way to begin solely 
with a one dimensional supersymmetrical theory as shown in (\ref{act1b}) 
or (\ref{act2b}) and determine which of the two four-dimensional theories 
could provide the starting point.  The way this is done is to note that for 
a one dimensional supersymmetrical theory, with at least four worldline 
SUSY charges, any action is also accompanied by an associated set 
of 4$\times$4 `L-matrices' and `R-matrices' \cite{G-1} as defined by 
the equations in (\ref{CM9}) 

All three chiral supermultiplets will have the same set of L-matrices 
and R-matrices as first derived before in the work of \cite{G-1}.  This same
work derived the L-matrices and R-matrices for the vector supermultiplet
also.  The work of \cite{permutadnk} noted each L-matrix and R-matrix 
can be expressed in terms of a `Boolean factor' denoted by $ ({\cal S
}^{ (\rI)})_i{}^\hl$ which appears via
\begin{equation}
 ({\rm L}_{\rI})_i{}^\hk ~=~ 
     ({\cal S}^{ (\rI)})_i{}^\hl\, ({\cal P}_{\! (\rI)})_\hl{}^\hk,
      \qquad \text{for each fixed }{\rm I}\, =\, 1,2,\dots,N.
\label{aas1}
\end{equation}
\begin{equation}
({\cal S}^{(\rI)})_i{}^\hl
=\begin{bmatrix}
(-1)^{b_1} & 0 & 0 & \cdots\\
0 & (-1)^{b_2} & 0 & \cdots\\
0 & 0 & (-1)^{b_3} & \cdots\\
\vdots & \vdots & \vdots & \ddots\\
\end{bmatrix}
\quad\iff\quad
\Big(\opR_{\rI} = \sum_{i=1}^{\rm d} b_i\, 2^{i-1}\Big)_b
 \label{binSign}
\end{equation}
(a diagonal matrix with real entries that squares to the identity) times an element 
of the permutation group $({\cal P}_{\! (\rI)})_\hl{}^\hk$.  The matrices above can 
be associated with a class of topological objects given the name of ``adinkras'' 
\cite{adinkra1} which are graphs that capture (with complete fidelity) the information 
in the matrices and nodal heights.   In fact, if the adinkras are regarded as graphs 
or networks, the permutation factor within the L-matrices and R-matrices are the 
`adjacency matrices' from graph theory.

The adinkras associated with the chiral and the vector supermultiplets, respectively, 
are shown below.  These are the graphs associated solely with the equations that 
appear in (A.1) and (A.4). 
$$
\vCent
{\setlength{\unitlength}{1mm}
\begin{picture}(-20,-140)
\put(-65,-45){\includegraphics[width=2in]{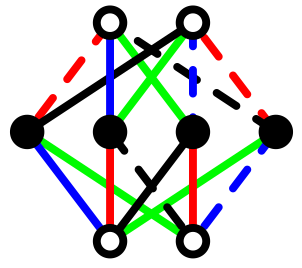}}
\put(-42,-51){(a.)}
\put(28,-51){(b.)}
\put(-50,-60){\bf {Figure \# 1: Adinkra graphs for the chiral (a.) and }}
\put(-23,-65){\bf {vector (b.) supermultiplets}}
\end{picture}}
$$ \newline
$$
\vCent
{\setlength{\unitlength}{1mm}
\begin{picture}(-20,-140)
\put(-1,-27){\includegraphics[width=2.55in]{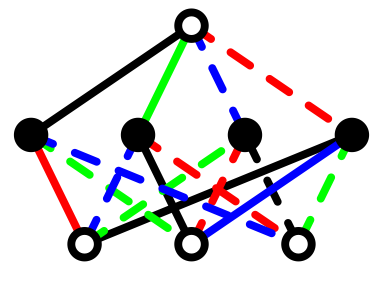}}
\end{picture}}
$$ 
\vskip1.6in
 On the other hand, the adinkra 
graphs associated with the valise equations of (\ref{CMv}) and (\ref{VMv}) 
are different and given in Figure \# 2.

Written solely in the form of valise adinkras or their associated 
matrices in Appendix B, it is not at all clear how the 
spin-bundle information to distinguish the chiral supermultiplet 
from the vector supermultiplet has been retained. The question 
becomes, ``What structure in the graphs or their associated matrices 
holographically stores the information about the distinction?'' 
$$
\vCent
{\setlength{\unitlength}{1mm}
\begin{picture}(-20,-140)
\put(-70,-29){\includegraphics[width=2.52in]{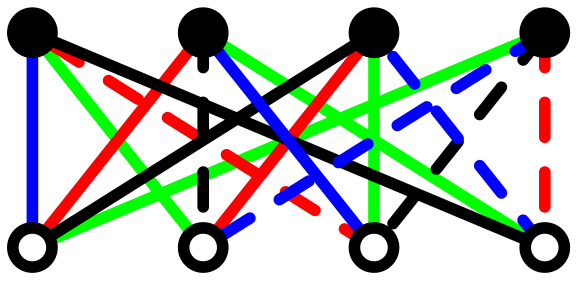}}
\put(01,-29.5){\includegraphics[width=2.35in]{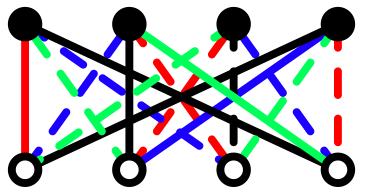}}
\put(-42,-38){(a.)}
\put(28,-38){(b.)}
\put(-56,-50){\bf {Figure \# 2: Adinkra graphs for the valise chiral (a.) }}
\put(-29,-55){\bf {and valise vector (b.) supermultiplets}}
\end{picture}}
$$ \newline
\vskip1.8in

\section{Adinkra Matrices and Information Conservation}

$~~~$ The work of \cite{permutadnk} has offered a proposal for identifying
such a mechanism: the information may be accessed via the elements of
the permutation group.   To see most transparently how the permutation 
group elements contain the information we are seeking, it is useful to 
describe these elements in terms of cycles\footnote{We acknowledge 
conversations with Kevin Iga who emphasized this point.}.

To show this approach clearly, we will give an explicit demonstration using the
L${}_{1}$ matrix of the chiral multiplet.
\be  \eqalign{
&\left( {\rm L}{}_{1}\right) {}_{i \, {\hat k}}   ~=~
\left[\begin{array}{cccc}
~1 & ~0 &  ~\, 0  &  ~0 \\
~0 & ~0 &  ~\, 0  &  -\, 1 \\
~0 & ~1 &  ~\,0  &  ~0 \\
~0 & ~0 &  -\, 1  &  ~0 \\
\end{array}\right]   ~=~ \left[\begin{array}{cccc}
~1 & ~0 &  ~~0  &  ~0 \\
~0 & ~ -\,1 &  ~~0  &  ~ 0 \\
~0 & ~ 0 &  ~1  &  ~0 \\
~0 & ~0 &  ~0  &  -\, 1 \\
\end{array}\right]    \,
\left[\begin{array}{cccc}
~1 & ~~0 &  ~~0  &  ~~0 \\
~0 & ~~0 &  ~~0  &  ~1 \\
~0 & ~~1 &  ~~0  &  ~0 \\
~0 & ~~0 &  ~ 1  &  ~0 \\
\end{array}\right]    \cr
&{~~~~~~~~~~~~~~~~~~~~~~~~~~~~~~}\left( {\rm L}{}_{1}\right) {}_{i \, {\hat k}}   
~=~  (10)_b \, \left[\begin{array}{cccc}
~1 & ~~0 &  ~~0  &  ~~0 \\
~0 & ~~0 &  ~~0  &  ~1 \\
~0 & ~~1 &  ~~0  &  ~0 \\
~0 & ~~0 &  ~ 1  &  ~0 \\
\end{array}\right]    ~~~,
}   \ee
where the Boolean factor $ (10)_b$ is defined according to the conventions of \cite{permutadnk}.
We next note that the element of the permutation group above obviously satisfies
\be
 \left[\begin{array}{cccc}
~1 & ~~0 &  ~~0  &  ~~0 \\
~0 & ~~0 &  ~~0  &  ~1 \\
~0 & ~~1 &  ~~0  &  ~0 \\
~0 & ~~0 &  ~ 1  &  ~0 \\
\end{array}\right]  \,  \left[\begin{array}{c}
~1  \\
~2  \\
~3 \\
~4 \\
\end{array}\right]   ~=~   \left[\begin{array}{c}
~1  \\
~4  \\
~2 \\
~3 \\
\end{array}\right] 
\ee
which implies
\be
1~ \to ~ 1 ~~,~~ 2~ \to ~ 4 ~~,~~ 3~ \to ~ 2 ~~,~~ 4~ \to ~ 3 ~~,
\ee
and this reveals the cycle $(2\, 4\,  3)$.  We then can write ${\rm L}_1 
= (10)_b \, (2\, 4\,  3)$.

Upon applying such considerations to all the L-matrices and the R-matrices, we find
\be  \eqalign{ {~}
&{\rm L}_1  \,=\, (10)_b \, (2\, 4\,  3) ~,~~ {\rm L}_2 \,=\, (12)_b \, (1\, 2\, 3)  
~,~~ {\rm L}_3  \,=\, (6)_b \, (1\, 3\, 4) ~\,~,~~\,  \,
{\rm L}_4   \,=\, (0)_b \, (1\, 4\, 2)~~,
~~  \cr
&{\rm R}_1  \,=\, (12)_b (2 \, 3 \, 4 )  \,\,,~\,  
{\rm R}_2  \,=\,  (9)_b \, (1\, 3\, 2)  ~~~,~\,  
{\rm R}_3   \,=\, (10)_b(1 \,  4\, 3) ~~,~~   
{\rm R}_4 \,=\, (0)_b \, (1\, 2 \, 4)
~\,~, } \label{LRmtrx0c}
\ee
for the chiral multiplet and
\be  \eqalign{ {~}
&{\rm L}_1 \,\,=\, (10)_b \, (1\, 2\, 4\,  3) ~~,~~ {\rm L}_2  \,=\, (12)_b \, (2\, 3)
\,~,~\,~ {\rm L}_3    \,=\, (0)_b \, (1\, 4) \,~,~\, 
{\rm L}_4  \,=\, (6)_b \,(1\, 3\, 4 \,2)  ~\,~,~~\,   
 \,~ \cr
 &{\rm R}_1 \,=\, (12)_b (1 \, 3 \, 4 \, 2 ) ~\,~,~~   {\rm R}_2  \,=\, (10)_b(2 \, 3) 
  ~~,~~  
 {\rm R}_3   \,=\, (0)_b \, (1\, 4) 
 ~,~~
 {\rm R}_4  \,=\, (13)_b \, (1\, 2\, 4 \, 3) 
~, } \label{LRmtrx0v}
\ee
for the vector multiplet.  

It is seen the chiral multiplet is associated with elements of the permutation 
group that include only three-elements cycles, while the vector multiplet is 
associated with elements of the permutation group that include only cycles
of `even length.'    As the formulation we use of the 4D, $\cal N$ = 4 supermultiplet 
possesses off-shell 4D, $\cal N$ = 1 supersymmetry, the distinction between 
the vector field and three scalars must be present for one SUSY charge in 
our subsequent discussion.  In fact, we will find that precisely this distinction
will be present for all four super charges.

From the adinkra networks shown in Figures \# 1 and \# 2,
the correlations between the link colors, cycles and L-matrices is 
shown in Table
\# 1 below.
$
\vCent
{\setlength{\unitlength}{1mm}
\begin{picture}(-20,-40)
\put(-110,-44){\bf {Table \# 1: Adinkra Link Color \& Cycles 
in L-matrices }}
\end{picture}}
$
\begin{table}[h] 
\vspace{0.2cm}
\begin{center}
\footnotesize
\begin{tabular}{|c|c|c|}  \hline 
$ {~} $ & $~~$CM$~~$ & $~~$VM$~~$ \\ \hline
$ BLUE $ & $ (243) $ & $ (1243) $ \\  \hline
$ RED $ & $ (123) $ & $ (23) $ \\  \hline
$ BLACK $ & $ (134) $ & $ (14) $ \\  \hline
$ GREEN $ & $ (142) $ & $ (1342) $ \\  \hline
\end{tabular}
\end{center}
\end{table} 
\noindent \vskip.05in  \noindent
The work of \cite{permutadnk} implies that the information about the
even vs.\ odd length cycles defines a `shadow' of the Hodge duality
that respects off-shell 4D, $\cal N$ = 1 supersymmetry.

\section{Valise 1D, $N$ = 16 Supermultiplet Formulation}

$~~~$ In this section we will present the new results of this work.   We
apply the result in (\ref{0brane}) and the field re-definitions in (\ref{ReDef})
to the action in (A.6) simultaneously to find
\be\eqalign{
{\mathcal{L}}~=~ &\frac{1}{2}(\partial_{\tau}A^{\cal I})(\partial_{\tau}A^{\cal I}) 
+ \frac{1}{2}(\partial_{\tau}B^{\cal I})(\partial_{\tau}B^{\cal I})
+\frac{1}{2}(\pa_{\tau}F^{\cal I})(\pa_{\tau}F^{\cal I})+\frac{1}{2}(\pa_{\tau}G^{\cal I})
(\pa_{\tau}G^{\cal I})
\cr
& +\frac{1}{2} (\partial_{\tau}A_i)  (\partial_{\tau}A_i) 
+\frac{1}{2}({\pa_{\tau} \rm d}) ({\pa_{\tau} \rm d}) 
+i\frac{1}{2}(\gamma \cdot {\cal T})^{ab}\psi_{a}^{\cal I}\partial_{\tau}\psi_{b
}^{\cal I} + i \frac{1}{2}(\gamma \cdot {\cal T})^{
cd}\lambda_{c} \pa_{\tau}  \lambda_{d} ~~~.
}  \label{n4actv} \ee

\be \eqalign{
{\rm D}_a A^{\cal J} ~=~&~ \psi_a^{\cal J}  ~~~~~~~~~~~\,~~~, ~~~
{\rm D}_a B^{\cal J} ~=~ ~i \, (\gamma^5){}_a{}^b \, \psi_b^{\cal J}  ~~~, \cr
{\rm D}_a F^{\cal J} ~=~&~ (\g \cdot {\cal T}){}_a{}^b \,  \psi_b^{\cal J}   
~~~, ~~~
{\rm D}_a G^{\cal J} ~=~ ~i \,(\gamma^5\g \cdot {\cal T}){}_a{}^b \, 
\psi_b^{\cal J}  ~~~,  \cr
{\rm D}_a \psi_b^{\cal J} ~=~ &~i\, (\g \cdot {\cal T}){}_{a \,b}\, \left( 
\partial_{\tau} A^{\cal J} \right) ~-~  (\gamma^5\g \cdot {\cal T}){}_{a 
\,b} \, \left( \partial_{\tau} B^{\cal J} \right) ~\cr
&-~ ~i \, C_{a\, b} \, \left( \partial_\tau
\,F^{\cal J}  \right)~+~ ~ (\gamma^5){}_{ a \, b} \, \left( \partial_\tau G^{\cal 
J} \right) ~~, \cr
} \label{chi0specif}
\ee 
\be \eqalign{
\D_a A_i & ~=~ (\gamma_i)_a{}^b \lambda_b  ~~~, ~~~ D_a {\rm d} 
 \, =\,  i (\gamma^5 \gamma \cdot {\cal T})_a{}^b \, \lambda_b 
 ~~~~~~~~~~~~~~, \cr
\D_a \lambda_b & ~=~ - \tfrac{i}{2}( [ \, \gamma \cdot {\cal T} ~,~ 
\gamma^i \, ])_{ab} \,  \left( \,  \pa_{\tau} A_i  \, \right)~+~ (\gamma^5
)_{ab} \,  \left( \, \pa_{\tau} {\rm d} \, \right)
~~~.  }    \label{VMvx}
\ee

However, for the SU(2) triplet supercovariant derivatives the realization
takes the forms
\be \eqalign{
{\rm D}_a^{\cal I} A^{\cal J} ~=~& \delta^{\cal {I \, J}}~ \l_a - \epsilon^{\cal {I 
\, J}}_{~~~{\cal K}} \,\psi_a^{\cal K}
~~~~~, ~~~~
{\rm D}_a^{\cal I} B^{\cal J} ~=~ ~i \, (\gamma^5){}_a{}^b \, \left[~\delta^{IJ}
~ \l_b + \epsilon^{\cal {I \, J}}_{~~~{\cal K}} \,\psi_b^{\cal K}~ \right]  ~~~, \cr
{\rm D}_a^{\cal I} F^{\cal J} ~=~& ~ (\g \cdot {\cal T}){}_a{}^b \,  
{\big [}~ \delta^{\cal {I \, J}}~ \l_b - \epsilon^{\cal {I \, J}}_{~~~{\cal K}} \,\psi_b^{\cal K}~
{\big ]}   ~~~, \cr
{\rm D}_a^{\cal I} G^{\cal J} ~=~& ~i \,(\gamma^5\g \cdot {\cal T}){}_a{}^b 
\,  {\big  [}~-\delta^{IJ}~ \l_b + \epsilon^{\cal {I \, J}}_{~~~{\cal K}} \,
\psi_b^{\cal K}~ {\big ]}  ~~~, \cr
{\rm D}_a^{\cal I} \psi_b^{\cal J} ~=~& \delta^{\cal {I \, J}} {\big [}~ i \, \fracm 12 ( [\, 
\g \cdot {\cal T} \, , \,  \gamma^{i} \,]){}_a{}_b \, (\,  \partial_{\tau}  
\, A{}_{i}   \, ) ~+~  (\gamma^5){
}_{a \,b} \,  \left( \pa_{\tau}  {\rm d}  \right) ~ {\big ]}  \cr
&+~\epsilon^{\cal {I \, J}}_{~~~{\cal K}} \,{\big [}~ i\, (\g \cdot {\cal T}){}_{a \,b}\,  
\left( \partial_{\tau} A^{\cal K} \right) ~+~ (\gamma^5\g \cdot {\cal 
T}){}_{a \,b} \, \left( \partial_{\tau} B^{\cal K} \right) ~\cr
&~~~~~~~~~~~-~ i \, C_{a\, b} \, \left( \partial_\tau F^{\cal K} \right)  
~-~ (\gamma^5){}_{ a \, b}\, \left( \partial_\tau G^{\cal K} \right)
~ {\big ]}  ~~, \cr
}\label{eq:chIspecif}
\ee
for the fields in the valise adinkra formulation of the three 
chiral supermultiplets  and 
\be \eqalign{
{\rm D}_a^{\cal I} \, A{}_{i} ~=~ & -(\gamma_i){}_a {}^b \,  
\psi_b^{\cal I}  ~~~, ~~~~
{\rm D}_a^{\cal I} \, {\rm d} ~=~ i \, (\gamma^5\g \cdot {\cal T}){}_a {}^b \, 
\,   \psi_b^{\cal I}  ~~~, \cr
{\rm D}_a^{\cal I} \l_b ~=~ & ~i\, (\g \cdot {\cal T}){}_{a \,b}\,  \left( 
\partial_\tau A^{\cal I} \right)~-~ ~ (\gamma^5\g \cdot {\cal T})
{}_{a \,b} \, \left( \partial_\tau B^{\cal I} \right) ~\cr
&-~i \, C_{a\, b} \, \left(  \partial_\tau F^{\cal I} \right)  ~-~ (\gamma^5)
{}_{ a \, b} \, \left( \partial_\tau G^{\cal I} \right)  ~~~.
} \label{VIspecifZZ}
\ee
for the fields of the valise adinkra formulation of the vector supermultiplet.
The equations in this section that involve the D-operators are clearly
of the same form as in (\ref{CM9}) with the indices now ranging as $\rm I$, 
$\rm J$, etc.  = 1, $\dots$, $16$; i, j, etc. = 1, $\dots$, 16; and $\hi$, $\hj$, 
etc. =  1, $\dots$, 16.

\section{Extracting 1D, $N$ = 16 Valise Adinkra Matrices}

$~~~$ With the results of the previous section in hand, we are now able to
extract the L-matrices and R-matrices of the 1D, $N$ = 16 adinkra matrices
associated with the discussion of the previous chapter.  In order to present
our results coherently, we use the following notation conventions that are
the most obvious appropriate generalizations of (\ref{CM9}).  We now
introduce the 1D covariant derivatives $\D^{[0]}{}_{\rm I}$ and $\D^{[{\cal I}
]}{}_{\rm I}$ to act as the holographic images of D${}_a$ and D${}_a{}^{\cal I}$.
Their realizations on the valise fields may be expressed in the forms
\be \eqalign{
\D^{[0]}{}_{\rm I} \Phi_i \= i\, (\L^{[0]}_{\rm I})_{i \hat{k}} \Psi_{\hat{k}} \qquad &,
\qquad \D^{[0]}{}_{\rm I} \Psi_{\hat{k}} \=  (\R^{[0]}_{\rm I})_{\hat{k} i} \, \left( 
\pa_{ \tau} \Phi_i  \right)
~~~,  \cr
\D^{[{\cal I}]}{}_{\rm I} \Phi_i \= i\, (\L^{[{\cal I}]}_{\rm I})_{i \hat{k}} \Psi_{\hat{k}} 
\qquad &, \qquad \D^{[{\cal I}]}{}_{\rm I} \Psi_{\hat{k}} \=  (\R^{[{\cal I}]}_{\rm 
I})_{\hat{k} i} \, \left( \pa_{ \tau}  \Phi_i  \right)  ~~~, 
} \label{CM9mod}
\ee
where above the bosonic and fermionic quantities $ \Phi_i$ and $ \Psi_{\hat{
k}}$ respectively take the forms of two sixteen component quantities
\be  \eqalign{
\Phi_i \,&=\,   \left(  A^1, \, B^1, \, F^1 , \, G^1 , \,   A^2, \, B^2, \, F^2 , \, G^2, \,  
A^3, \, B^3, \, F^3 , \, G^3 , \, {\vec A}, \, {\rm d}   \right)  ~~~,
\cr
\Psi_{\hat{k}} \,&=\,  - i {\Big( } \psi^1{}_1 , \, \psi^1{}_2 , \, \psi^1{}_3 , \, \psi^1
{}_4 , \,  \psi^2{}_1 , \, \psi^2{}_2 , \, \psi^2{}_3 , \, \psi^2{}_4 , \, \psi^3{}_1 , \, 
\psi^3{}_2 , \, \psi^3{}_3 , \, \psi^3{}_4 , \,  \l{}_1 , \, \l{}_2 , \, \l{}_3 , \, \l{}_4 
{ \Big )} ~,~~~
}   \label{16Vals}
\ee
(where $\vec A$ corresponds to the spatial components of the gauge field) 
as is appropriate in the context of this chapter.

Explicitly we find for the $(\L^{[0]}_{\rm I})_{i \hat{k}}$ matrices
$$
\left( {\rm L}{}_{1}^{[0]}\right) {}_{i   {\hat k}}    ~=~ \left[\begin{array}{cccc}
  (10)_b   (243) &  0 &   0  &   0 \\
 0 &   (10)_b    (243) &   0  &   0 \\
 0 &  0 &    (10)_b   (243)  &   0 \\
 0 &  0 &   0  &    (10)_b   (1243)\\
\end{array}\right]   {,}  {}
$$
$$
\left( {\rm L}{}_{2}^{[0]}\right) {}_{i   {\hat k}}    ~=~ \left[\begin{array}{cccc}
  (12)_b   (123) &  0 &   0  &   0 \\
 0 &   (12)_b   (123) &   0  &   0 \\
 0 &  0 &     (12)_b   (123)  &   0 \\
 0 &  0 &   0  &    (4)_b   (23)\\
\end{array}\right]   { ,}  {}
$$
$$
\left( {\rm L}{}_{3}^{[0]}\right) {}_{i   {\hat k}}    ~=~ \left[\begin{array}{cccc}
  (6)_b   (134) &  0 &   0  &   0 \\
 0 &   (6)_b   (134)  &   0  &   0 \\
 0 &  0 &    (6)_b   (134)   &   0 \\
 0 &  0 &   0  &    (0)_b   (14)\\
\end{array}\right]  {,}  {}
$$
\be
\left( {\rm L}{}_{4}^{[0]}\right) {}_{i   {\hat k}}    ~=~ \left[\begin{array}{cccc}
  (0)_b   (142) &  0 &   0  &   0 \\
 0 &   (0)_b   (142)  &   0  &   0 \\
 0 &  0 &    (0)_b   (142)   &   0 \\
 0 &  0 &   0  &    (6)_b   (1342)\\
\end{array}\right]  {,}  {} \label{vN4Lsingl}
\ee
and these L-matrices are simply reaffirming relations of colors to eight distinct 
cycles seen before in chapter three.  In a similar manner  the $(\R^{[0]}_{\rm 
I})_{\hat{k} \, i}$ matrices take the forms
$$
\left( {\rm R}{}_{1}^{[0]}\right) {}_{{\hat k} \, i}   ~=~
\left[\begin{array}{cccc}
(12)_b (234) & 0 & 0 & 0 \\
0 & (12)_b (234) & 0 & 0 \\
0 & 0 & (12)_b (234) & 0 \\
0 & 0 & 0 & (12)_b (1342)
\end{array}\right]
$$
$$
\left( {\rm R}{}_{2}^{[0]}\right) {}_{{\hat k} \, i}   ~=~
\left[\begin{array}{cccc}
(9)_b(132) & 0 & 0 & 0 \\
0 & (9)_b(132) & 0 & 0 \\
0 & 0 & (9)_b(132) & 0 \\
0 & 0 & 0 & (10)_b(23)
\end{array}\right]
$$
$$
\left( {\rm R}{}_{3}^{[0]}\right) {}_{{\hat k} \, i}   ~=~
\left[\begin{array}{cccc}
(10)_b(143) & 0 & 0 & 0 \\
0 & (10)_b(143) & 0 & 0 \\
0 & 0 & (10)_b(143) & 0 \\
0 & 0 & 0 & (0)_b(14)
\end{array}\right]
$$
\be
\left( {\rm R}{}_{4}^{[0]}\right) {}_{{\hat k} \, i}   ~=~
\left[\begin{array}{cccc}
(0)_b (124) & 0 & 0 & 0 \\
0 & (0)_b(124) & 0 & 0 \\
0 & 0 & (0)_b(124) & 0 \\
0 & 0 & 0 & (9)_b (1243)
\end{array}\right]
 \label{vN4Rsingl}
\ee
in the basis defined by (\ref{16Vals}).

This brings us to the explicit results for the triplet L-matrices and R-matrices.

One of the most obvious features about these is that the holographical
mechanism for conserving the spin-bundle information of the four dimensional
related theory is still present in the 1D, $N$ = 16 valise formulation!  The 
explicit way this occurs is by making three observations: 
\newline (a.) examination of the matrices in (\ref{vN4Lsingl}) 
and (\ref{vN4Rsingl}) shows a 3:1 ratio of odd-length \newline \indent $\,$ 
cycles/even-length cycle within each L-matrix and R-matrix, 
\newline (b.) examination of the matrices in Appendix C continues
to show a 3:1 ratio of \newline \indent $\,$
three odd-length cycles/even-length cycle within each L-matrix and R-matrix
 \newline \indent $\,$
cycle, and
\newline (c.) examination of the matrices in Appendix C
shows that only the {\em {same}} eight \newline \indent $\,$ cycles within the 
permutation group appear within the triplet L-matrices as \newline \indent $\,$ 
do appear in the singlet L-matrices.

\noindent
The first of these observations was expected as it follows from the fact 
that the first supersymmetry generated by the singlet D${}_a$-operator obviously 
acts on three 1D, $N$ = 4 chiral multiplet adinkras and one 1D, $N$ = 4 vector 
multiplet adinkra. 

The second and third observations, however, are striking evidence that the 
mechanism  of using the lengths of cycles of the permutation elements embedded 
with the L-matrices and R-matrices apparently continues to work as the 
transformation laws associated with the hologram of the triplet D${}_a^{\cal 
I}$-operators possesses the same property and there was {\em {no}} {\em {a}} 
{\em {priori}} reason to expect this conservation of information among all four 
supercharges.   Furthermore, it is seen that in all four sets of L-matrices 
associated with the supercharges, the same 3:1 ratio of odd-length 
three-cycles/even cycles is present. 

We are thus fortified in our assertion that the 1D, $N$ = 16 SUSY quantum
mechanical model described by the equations in chapter five constitutes
a SUSY hologram of the 4D, $\cal N$ = 4 model described in chapter
four.  

But is it an off-shell valise adinkra hologram?

\section{Previous Results About Off-Shell vs. On-Shell}

$~~~$ In order to answer this question, it is useful to both review some 
previous work \cite{G-1} that can be used as a foundation upon which an
expanded discussion can be built.  In this chapter we use the on-shell
versus off-shell valise adinkra formulations of the 4D, $\cal N$ = 1 chiral 
scalar and the vector supermultiplets as our jumping off points.

From the perspective of valise adinkra formulations, L-matrices and
R-matrices continue to exist for on-shell theories but with the main distinctions:
\newline (a.) the i, j, etc. indices have a range according to 1, 
$\dots$, d${}_L$, the $\hi$, $\hj$, etc.  \newline \indent $\,$ 
indices have a range according to 1, $\dots$, d${}_R$, where 
d${}_L$ may be distinctly \newline \indent $\,$ different  
from d${}_R$, and
\newline (b.) the algebras for the L-matrices and
R-matrices are changed to:
\be \eqalign{
 (\,{\rm L}_\rI\,)_i{}^\hj\>(\,{\rm R}_\rJ\,)_\hj{}^k + (\,{\rm L}_\rJ\,)_i{}^\hj\>(\,{\rm 
 R}_\rI\,)_\hj{}^k &= 2\,\delta_{\rI\rJ}\,\delta_i{}^k ~+~ \left( {\Delta }^{\rm L}_{\rI \, 
 \rJ} \right)_i{}^k 
 ~~,\cr
 (\,{\rm R}_\rJ\,)_\hi{}^j\>(\, {\rm L}_\rI\,)_j{}^\hk + (\,{\rm R}_\rI\,)_\hi{}^j\>(\,{\rm 
 L}_\rJ\,)_j{}^\hk
  &= 2\,\delta_{\rI\rJ}\,\delta_\hi{}^\hk  ~+~ \left( {\Delta}^{\rm R}_{\rI \, \rJ} 
  \right)_\hi{}^\hk~~.
}  \label{GarDNAlg1ON}
 \ee
The quantities $\left( {\Delta }^{\rm L}_{\rI \,  \rJ} \right)_i{}^k $ and
$\left( {\Delta}^{\rm R}_{\rI \, \rJ}  \right)_\hi{}^\hk$
respectively measure the non-closure of the SUSY algebra on
the bosons and fermions of the supermultiplet.

\subsection{On-Shell Chiral Supermultiplet}

$~~~$ The fields $\left( A, \, B, \,  \psi_a  \right)$ of the on-shell valise version 
of the 4D, $\cal N$ = 1 chiral are shown in Figure
\# 3 below.
$$ 
\vCent
{\setlength{\unitlength}{1mm}
\begin{picture}(-20,-140)
\put(-30,-31){\includegraphics[width=2.3in]{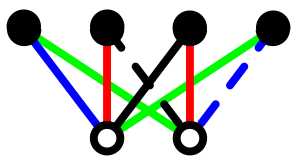}}
\put(-60,-38){\bf {Figure \# 3: Adinkra for On-shell Chiral Supermultiplet }}
\label{onCMvv}
\end{picture}}
$$ \vskip1.3in
\noindent
which correspond to the D-algebra equations
\be \eqalign{
{\rm D}_a A ~&=~ \psi_a  ~~~, ~~~
{\rm D}_a B ~=~ i \, (\gamma^5){}_a{}^b \, \psi_b  ~~~, \cr
{\rm D}_a \psi_b ~&=~ i\, (\gamma \cdot {\cal T}){}_{a \,b}\,  \partial_{\tau} A 
~-~  (\gamma^5\gamma \cdot {\cal T}){}_{a \,b} \, \partial_{\tau} B   ~~~.
} \label{chi3z}
\ee
Using (\ref{chi3z}), calculations yield the follow super-commutator algebra
\be \eqalign{  {~~~~~}
\{ ~ {\rm D}_a  \,,\,  {\rm D}_b ~\} \, A 
~&=~  i\, 2 \, (\gamma \cdot {\cal T}){}_{a \,b}\,  \partial_{\tau} \,  A ~~~, ~~~
\{ ~ {\rm D}_a  \,,\,  {\rm D}_b ~\} \, B 
~=~  i\, 2 \, (\gamma \cdot {\cal T}){}_{a \,b}\,  \partial_{\tau} \, B ~~~, \cr
\{ ~ {\rm D}_a  \,,\,  {\rm D}_b ~\} \, \psi{}_{c}  
~&=~  i\, 2 \, (\gamma \cdot {\cal T}){}_{a \,b}\,  \partial_{\tau} \,  \psi{}_{c}    
~-~   i \, (\gamma^{\mu}){}_{a \,b}\, (\gamma_\mu
\gamma \cdot {\cal T}){}_c{}^d  \partial_{\tau} \,  \psi{}_{d}     ~~~.
} \label{chi4}
\ee
The first two of these equations have the expected form for a supersymmetry
algebra, but the third term immediately above can be re-expressed as
\be \eqalign{
\{ ~ {\rm D}_a  \,,\,  {\rm D}_b ~\} \, \psi{}_{c}  
~&=~  i\, 2 \, (\gamma \cdot {\cal T}){}_{a \,b}\,  \partial_{\tau} \,  \psi{}_{c}    
~+~  i\, 2 \, (\gamma^{\mu}){}_{a \,b}\, (\gamma_\mu){}_c{}^d 
{\cal K}{}_{d} (\psi) ~~,~~  \cr
 {\cal K}{}_{c} (\psi) ~&=~ - \, \fracm 12 \,  (
\gamma \cdot {\cal T}){}_c{}^d  \partial_{\tau} \,  \psi{}_{d}     ~~~,
}  \label{chi5}
\ee
where $ {\cal K}{}_ c$ measures the `non-closure' of the algebra.   From the
adinkra in Figure \# 3, we define the (2 $\times$ 1) bosonic ``field vector'' 
and (4 $\times$ 1) fermionic ``field vector'' 
\be  \eqalign{
\Phi_i \,&=\,   \left(  A, \, B \,   \right)  ~~~,  ~~~
\Psi_{\hat{k}} \,=\,  - i \left(  \psi{}_1 , \, \psi{}_2 , \, \psi{}_3 , \, \psi
{}_4   \right) ~,~~~
}   \label{ChiVal}
\ee
as appropriate for such the adinkra shown in (\ref{onCMvv}).  This permits 
us to obtain the following L-matrices and R-matrices.
$$
\left( {\rm L}{}_{1}\right) {}_{i \, {\hat k}}   ~=~
\left[\begin{array}{cccc}
~1 & ~~0 &  ~~0  &  ~~0 \\
~0 & ~~0 &  ~~0  &  ~-\, 1 \\
\end{array}\right] ~~~,~~~
\left( {\rm L}{}_{2}\right) {}_{i \, {\hat k}}   ~=~
\left[\begin{array}{cccc}
~0 & ~~1 &  ~~0  &  ~ \, \, 0 \\
~0 & ~~ 0 &  ~~1  &  ~~0 \\
\end{array}\right]  ~~~,
$$
\be
\left( {\rm L}{}_{3}\right) {}_{i \, {\hat k}}   ~=~
\left[\begin{array}{cccc}
~0 & ~~0 &  ~~1  &  ~~0 \\
~0 & ~- \, 1 &  ~~0  &  ~~0 \\
\end{array}\right] ~~~,~~~
\left( {\rm L}{}_{4}\right) {}_{i \, {\hat k}}   ~=~
\left[\begin{array}{cccc}
~0 & ~~0 &  ~~0  &  ~ \, \, 1 \\
~1 & ~~ 0 &  ~~0  &  ~~0 \\
\end{array}\right]  ~~~,
 \label{chiD0N}
\ee
$$
\left( {\rm R}{}_{1}\right) {}_{i \, {\hat k}}   ~=~
\left[\begin{array}{cc}
~1 & ~~0  \\
~0 & ~~0  \\
~0 & ~~0  \\
~0 & -1  \\
\end{array}\right] ~~~,~~~
\left( {\rm R}{}_{2}\right) {}_{i \, {\hat k}}   ~=~
\left[\begin{array}{cc}
0 & ~~0  \\
~1 & ~~ 0  \\
~0 & ~~ 1  \\
~ 0 & ~~~0  \\
\end{array}\right]  ~~~,
$$
\be
\left( {\rm R}{}_{3}\right) {}_{i \, {\hat k}}   ~=~
\left[\begin{array}{cc}
~0 & ~~0  \\
~0 & - \, 1  \\
~1 & ~~0  \\
~0 & ~~0  \\
\end{array}\right] ~~~,~~~
\left( {\rm R}{}_{4}\right) {}_{i \, {\hat k}}   ~=~
\left[\begin{array}{cc}
~0 & ~~1  \\
~0 & ~~ 0  \\
~0 & ~~ 0  \\
~ 1 & ~~~0   \\
\end{array}\right]  ~~.
 \label{chiD0O}
\ee
Given the matrices in (\ref{chiD0N}) and (\ref{chiD0O}) we find the following relations hold
\be \eqalign{
{~~~~~}
 (\,{\rm L}_\rI\,)_i{}^\hj\>(\,{\rm R}_\rJ\,)_\hj{}^k + (\,{\rm L}_\rJ\,)_i{}^\hj\>(\,{\rm 
 R}_\rI\,)_\hj{}^k &= 2\,\delta_{\rI\rJ}\,\delta_i{}^k  ~~,\cr
(\,{\rm R}_\rJ\,)_\hi{}^j\>(\,{\rm L}_\rI\,)_j{}^\hk + (\,{\rm R}_\rI\,
)_\hi{}^j\>(\,{\rm L}_\rJ\,)_j{}^\hk
&=~  \delta{}_{\rI\rJ} \, ({\bf I})_\hi{}^\hk 
 ~+~  \ [\, {\vec {\a}} {\b}^1 \, ]_{\rI\rJ}\,  \cdot \,
(\, {\vec {\a}} {\b}^1 \, )_\hi{}^\hk   
~~.
}  \label{NotGDNAlg1}
\ee

\subsection{On-Shell Vector Supermultiplet}

$~~~$ In an on-shell vector supermultiplet theory, we have the fields $(A{}_i , \, \l_a )$
may be represented by an adinkra of the form
$$
\vCent
{\setlength{\unitlength}{1mm}
\begin{picture}(-60,-100)
\put(-30,-21){\includegraphics[width=2.35in]{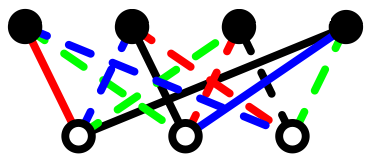}}
\put(-60,-27){\bf {Figure \# 4: Adinkra for On-shell Chiral Supermultiplet }}
\label{onVMvv}
\end{picture}}
$$ \vskip1in \noindent
which may be chosen to satisfy the equations
\be \eqalign{
{\rm D}_a \, A{}_{i} ~&=~  (\gamma_i){}_a {}^b \,  \l_b  ~~~, \cr
{\rm D}_a \l_b ~&=~   - \,i \, \fracm 12 ( [\, \gamma \cdot {\cal T} \, , \,  \gamma^{i} 
\,]){}_a{}_b \, (\,  \partial_{\tau}  \, A{}_{i} 
\, )  ~~~. \cr
} \label{V4}
\ee

From the adinkra above, we define the (3 $\times$ 1) bosonic ``field vector'' 
and (4 $\times$ 1) fermionic ``field vector'' 
\be  \eqalign{
\Phi_i \,&=\,   \left(  A_1, \, A_2,  \, A_3 \,   \right)  ~~~,  ~~~
\Psi_{\hat{k}} \,=\,  - i \left(  \lambda{}_1 , \, \lambda{}_2 , \, \lambda{}_3 , \, \lambda
{}_4   \right) ~,~~~
}   \label{vecVal}
\ee
as appropriate for such the adinkra shown in Figure \# 4 and once again we 
calculate the anti-commutator as realized on the remaining fields to find
\be \eqalign{
{~~~~~~~}
\{ ~ {\rm D}_a  \,,\,  {\rm D}_b ~\} \, A{}_{i}  
~&=~  i\, 2 \, (\gamma \cdot {\cal T}){}_{a \,b}\,  \partial_{\tau} \, A{}_{i}  
~~,    \cr
\{ ~ {\rm D}_a  \,,\,  {\rm D}_b ~\} \,\l{}_c
~&=~  i\, 2 \, (\gamma \cdot {\cal T}){}_{a \,b}\,  \partial_{\tau}  \,  \l{}_c   ~-~ i \,
\fracm 12 \,  (\gamma^\mu){}_{a \,b}\,  (\gamma_\mu  
\gamma \cdot {\cal T}){}_c{}^d \, \partial_{\tau}  \,  \l{}_d   \cr
&~~~~+~ i \, \fracm 1{16} \,  ([ \, \gamma^\a \, , \, \gamma^\b \,]){}_{a \,b}\,  
( [ \, \gamma_\a \, , \, \gamma_\b \,] \gamma \cdot {\cal T}){}_c{}^d \,   \partial_{\tau}  \,  \l{}_d   
~~.
} \label{V5}
\ee
The final equation of (\ref{V5}) shows the presence of {\em {two}} non-closure 
terms.  We may rewrite the final line as
\be \eqalign{
\{ ~ {\rm D}_a  \,,\,  {\rm D}_b ~\} \,\l{}_c
~&=~  i\, 2 \, (\gamma \cdot {\cal T}){}_{a \,b}\,  \partial_{\tau} \,  \l{}_c   ~+~ i 2 \,
 (\gamma^\mu){}_{a \,b}\,  (\gamma_\mu  ){}_c{}^d \,  {\Hat K}{}_d(\l)   \cr
&~~~~-~ i \, \fracm 1{4} \,  ([ \, \gamma^\a \, , \, \gamma^\b \,]){}_{a \,b}\,  
( [ \, \gamma_\a \, , \, \gamma_\b \,] ){}_c{}^d \,  {\Hat K}{}_d(\l)  ~~, \cr
{\Hat K}{}_c(\l) ~&\equiv ~  -\, \fracm 14 (\gamma \cdot {\cal T}){}_c{}^d \,   
\partial_{\tau} \, \l{}_d  ~~,
} \label{V6}
\ee
where the non-closure term $  {\Hat K}{}_c(\l)$ is introduced.  Given the 
bosonic field vectors, the fermionic field vector, and the valise adinkra
in Figure \# 4, we find the following L-matrices and R-matrices below.
$$
\left( {\rm L}{}_{1}\right) {}_{i \, {\hat k}}   ~=~
\left[\begin{array}{cccc}
~0 & ~1 &  ~ 0  &  ~ 0 \\
~0 & ~0 &  ~0  &  -\,1 \\
~1 & ~0 &  ~ 0  &  ~0 \\
\end{array}\right] ~~~,~~~
\left( {\rm L}{}_{2}\right) {}_{i \, {\hat k}}   ~=~
\left[\begin{array}{cccc}
~1 & ~ 0 &  ~0  &  ~ 0 \\
~0 & ~ 0 &  ~1  &  ~ 0 \\
 ~0 & - \, 1 &  ~0  &   ~ 0 \\
\end{array}\right]  ~~~,
$$
\be {~~~~}
\left( {\rm L}{}_{3}\right) {}_{i \, {\hat k}}   ~=~
\left[\begin{array}{cccc}
~0 & ~0 &  ~ 0  &  ~ 1 \\
~0 & ~1 &  ~0  &   ~0 \\
~0 & ~0 &  ~ 1  &  ~0 \\
\end{array}\right] ~~~~~~,~~~
\left( {\rm L}{}_{4}\right) {}_{i \, {\hat k}}   ~=~
\left[\begin{array}{cccc}
~0 & ~0 &  ~1  &  ~ 0 \\
-\,1 & ~ 0 &  ~0  &  ~ 0 \\
 ~0 & ~0 &  ~0  &   - \, 1 \\
\end{array}\right]  ~~~,
\label{V1D0H}
\ee

$$
\left( {\rm R}{}_{1}\right) {}_{ {\hat k} \, i}   ~=~
\left[\begin{array}{ccc}
~0 & ~0 &  ~1   \\
~1 & ~ 0 &  ~0   \\
 ~0 & ~0 &  ~0   \\
 ~0 & -1 &  ~0   \\
\end{array}\right] ~~~~~~,~~~
\left( {\rm R}{}_{2}\right) {}_{ {\hat k} \, i}   ~=~
\left[\begin{array}{ccc}
~1 & ~0 &  ~0   \\
~0 & ~ 0 &  -1   \\
 ~0 & ~1 &  ~0   \\
 ~0 & ~ 0 &  ~0   \\
\end{array}\right]  ~~~,
$$

\be
\left( {\rm R}{}_{3}\right) {}_{ {\hat k} \, i}   ~=~
\left[\begin{array}{ccc}
~0 & ~0 &  ~0   \\
~0 & ~ 1 &  ~0   \\
 ~0 & ~0 &  ~1   \\
 ~1 & ~ 0 &  ~0   \\
\end{array}\right] ~~~~~~,~~~
\left( {\rm R}{}_{4}\right) {}_{ {\hat k} \, i}   ~=~
\left[\begin{array}{ccc}
~0 & -\,1 &  ~1   \\
~0 & ~ 0 &  ~0   \\
 ~1 & ~0 &  ~0   \\
 ~0 & ~ 0 &  -\,1   \\
\end{array}\right]  ~~~,
\label{V1D0HO}   
\ee
Given the matrices in (\ref{V1D0H}) and (\ref{V1D0HO}) we find the following 
relations hold
\be \eqalign{
{~~~~~}
 (\,{\rm L}_\rI\,)_i{}^\hj\>(\,{\rm R}_\rJ\,)_\hj{}^k + (\,{\rm L}_\rJ\,)_i{}^\hj\>(\,{\rm 
 R}_\rI\,)_\hj{}^k &= 2\,\delta_{\rI\rJ}\,\delta_i{}^k  ~~,\cr
(\,{\rm R}_\rJ\,)_\hi{}^j\>(\,{\rm L}_\rI\,)_j{}^\hk + (\,{\rm R}_\rI\,)_\hi{}^j\>(\,{\rm 
L}_\rJ\,)_j{}^\hk &=~ \fracm 32 \, {\delta}_{\rI\rJ}\,  (\,  {\bf I}_4  \,)_\hi{}^\hk 
~-~ \fracm 12 \,  [\,  {\vec \a} \,  \b^2  \, ]_{\rI\rJ}\, \cdot \,  (\,  {\vec \a} \,  \b^2 
\, )_\hi{}^\hk  \cr
&~~~~+~ \fracm 12 \,  [\,   {\vec \a} \,  \b^1 \, ]_{\rI\rJ}\, \cdot \,  (\,  {\vec \a} \, 
\b^1   \, )_\hi{}^\hk  \cr
&~~~~+~ \fracm 12 \,  [\,   {\vec \a} \,  \b^3 \, ]_{\rI\rJ}\, \cdot \,  (\,  {\vec \a} \, 
\b^3   \, )_\hi{}^\hk  ~~.
}  \label{NotGDNAlg4}
 \ee

The results in (\ref{chi4}), (\ref{NotGDNAlg1}), (\ref{V5}), and (\ref{NotGDNAlg4})
once more are beautiful examples of SUSY holography.  In the cases of
 (\ref{chi4}), and (\ref{V5}) the SUSY commutator algebra closes on the
 bosons.  In the corresponding algebra of the L-matrices and R-matrices,
 the quantity $\left( {\Delta }^{\rm L}_{\rI\,\rJ} \right)_i{}^k$ is identically zero.
 
However, a careful examination of $\left( {\Delta}^{\rm R}_{\rI \, \rJ}\right)_\hi{}^\hk$
in each of the respective cases (\ref{NotGDNAlg1}) and (\ref{NotGDNAlg4})
reveals an even more striking exhibition of SUSY holography.  

In the case of (\ref{chi5}), if we look at the non-closure terms, it is seen that there 
appear four linearly independent matrices $(\g^{\mu})_{a b}$ on the right hand
side of the equation.  In a similar manner, if we look at  $\left( {\Delta}^{\rm 
R}_{\rI \, \rJ}\right)_\hi{}^\hk$ (as defined by the second line of (\ref{NotGDNAlg1})), 
we see there appear precisely four independent matrices ${\delta}_{\rI \, \rJ}$,
and $({\vec {\a}} {\b}^1)_{\rI \, \rJ}$.

In the case of (\ref{V6}), if we look at the non-closure terms, it is seen that there 
appear ten linearly independent matrices $(\g^{\mu})_{a b}$, and $([ \, \gamma_\a \, , \, 
\gamma_\b \,] )_{a b}$ on the right hand side of the equation.  In a similar manner, if 
we look at  $\left( {\Delta}^{\rm R}_{\rI \, \rJ}\right)_\hi{}^\hk$ (as defined by the 
second line of (\ref{NotGDNAlg4})), we see there appear precisely ten 
independent matrices  ${\delta}_{\rI \, \rJ}$, $({\vec \a}\b^2)_{\rI \, \rJ}$,  $(
{\vec \a}\b^1)_{\rI \, \rJ}$, $({\vec \a}\b^3)_{\rI \, \rJ}$
 on the right hand side of the equation. 
 
Finally, by looking at the results in (\ref{chi5}), and (\ref{V6}), it is seen that
the concept of an ``off-shell central charge'' collapses for one dimensional 
SUSY theories!  In higher dimensions imposing the conditions $ {\cal K}{}_{c} 
(\psi)$ = 0 or ${\Hat K}{}_{c} (\lambda)$ = 0 is equivalent to imposition of
a equation of motion restrictions on fermions.  However, for a one dimensional
SUSY theory, this is also equivalent to demanding that all fermionic fields
are constants and by consistency all bosonic field can be at most linear
functions of $\tau$.

\subsection{The 0-Brane Reduced Algebra}\label{CCCVReductionNequals4Algebra}

$~~~$ Here we present the form of the super-commutator algebra
of the supercovariant derivatives whose realizations are given in 
\ref{chi0specif} -  \ref{VIspecifZZ}.
\be\label{eq:IJtermsspecific1} 
\eqalign{
\{{\rm D}_a^{\cal I}, {\rm D}_b^{\cal J}\} A^{\cal K} ~=~ &i\, 2 \delta^{\cal 
{I \, J}} ( \g \cdot {\cal T})_{ab}  \pa_{\tau} A^{\cal K} - 2 \epsilon^{\cal {I \, J 
\, K}} (\g^5)_{ab} \,  \pa_{\tau} {\rm d} 
{~~~~~~~~~~~~~~~~~~~~~~~~~~~~} \cr
&-2 \k^{\cal {I \, J \, K \, M}}[i C_{ab}  \pa_{\tau} F^{\cal M} + (\g^5)_{ab}
 \pa_{\tau} G^{\cal M}]] ~~~,  \cr
\{{\rm D}_a^{\cal I}, {\rm D}_b^{\cal J}\} B^{\cal K} ~=~ &i\, 2 \delta^{\cal 
{I \, J}} ( \g \cdot {\cal T})_{ab}  \pa_{\tau} B^{\cal K} + 2~i \epsilon^{\cal {I \, 
J \, K}} C_{ab} \,  \pa_{\tau} {\rm d} ~~~,  \cr
\{{\rm D}_a^{\cal I}, {\rm D}_b^{\cal J}\} F^{\cal K} ~=~ &i\, 2 \delta^{\cal 
{I \, J}} ( \g \cdot {\cal T})_{ab}  \pa_{\tau} F^{\cal K} + 2 \epsilon^{\cal {I \, J 
\, K}} (\g^5\g \cdot {\cal T})_{ab} \pa_{\tau} {\rm d}  \cr
&+ 2 \k^{\cal {I \, J \, K \, M}}[ -i C_{ab}\partial_{\tau} A^{\cal M} + (\g^5\g \cdot {\cal T}
)_{ab} \pa_{\tau} G^{\cal M}] ~~~,  \cr
\{{\rm D}_a^{\cal I}, {\rm D}_b^{\cal J}\} G^{\cal K} ~=~ &i\, 2 \delta^{\cal 
{I \, J}} ( \g \cdot {\cal T})_{ab}  \pa_{\tau} G^{\cal K} + 2 \epsilon^{\cal {I \, J 
\, K}} (\g^5\g^{i})_{ab} \, \partial_{\tau} A_{i} \cr
 &- 2 \k^{\cal {I \, J \, K \, M}}[(\g^5)_{ab}\partial_{\tau} A^{\cal M}  + (\g^5
 \g \cdot {\cal T})_{ab} \pa_{\tau} F^{\cal M}] ~~~, \cr
\{{\rm D}_a^{\cal I}, {\rm D}_b^{\cal J}\} {\rm d} ~=~ &i\, 2 \delta^{\cal 
{I \, J}} ( \g \cdot {\cal T})_{ab}  \pa_{\tau} {\rm d} \cr
&+ 2 \epsilon^{\cal {I \, J \, K}} \left[ \, (\g^5)_{ab}\partial_{\tau} A^{\cal K} - 
i  C_{ab}\partial_{\tau} B^{\cal K} + (\g^5\g \cdot {\cal T})_{ab} \pa_{\tau} 
F^{\cal K} \, \right] ~~~, \cr
\{{\rm D}_a^{\cal I}, {\rm D}_b^{\cal J}\} A_i ~=~ & i\, 2 \delta^{\cal {
I \, J}} ( \g \cdot {\cal T})_{ab}  \pa_{\tau} A_i - 2 \epsilon^{\cal {I 
\, J \, K}} \,  (\g^5\g_i)_{ab}  \pa_{\tau} G^{\cal K} ~~~,  \cr
\{{\rm D}_a^{\cal I}, {\rm D}_b^{\cal J}\} \lambda_c ~=~ & i\, 2 \delta^{
\cal {I \, J}} ( \g \cdot {\cal T})_{ab}  \pa_{\tau} \lambda_c - i \epsilon^{\cal {
I \, J \, K}} [\, C_{ab} (\g \cdot {\cal T})_c^{~d} - (\g^5)_{ab}(\g^5
\g \cdot {\cal T})_c{}^d  \cr 
&~~~~~~~~~~~~~~~~~~~~~~~~~~~~~~~~~-(\g^5
\g^\nu)_{a \,b}(\g^5\g_\nu\g \cdot {\cal T})_c{
}^d]  \, \pa_{\tau} \psi_d^{\cal K}  ~~~, \cr
\{{\rm D}_a^{\cal I}, {\rm D}_b^{\cal J}\} \psi_c^{\cal K} ~=~ & i\, 2 
\delta^{\cal {I \, J}} ( \g \cdot {\cal T})_{ab}  \pa_{\tau} \psi_c^{\cal K} + i 
\epsilon^{\cal {I \, J \, K}} [\, C_{ab} (\g \cdot {\cal T})_c^{~d} - (\g^5)_{ab}(\g^5
\g \cdot {\cal T})_c{}^d  \cr 
&~~~~~~~~~~~~~~~~~~~~~~~~~~~~~~~~~~~~~-(\g^5
\g^\nu)_{a \, b}(\g^5\g_\nu\g \cdot {\cal T})_c{
}^d] \, \pa_{\tau} \lambda_d   \cr  
&~~~~~~~~~~~~~~~~~~- i \k^{\cal {I \, J \, K \, M}}[\, C_{ab}(\g \cdot {\cal T}
)_c{}^d + (\g^5)_{ab}(\g^5\g \cdot {\cal T})_c{}^d   \cr 
&~~~~~~~~~~~~~~~~~~~~~~~~~~~~~~~~+(\g^5\g^\nu)_{
ab}\, (\g^5\g_\nu\g \cdot {\cal T})_c{}^d] \,  \pa_{\tau} 
\psi_d^{\cal M}  ~~~.
}\ee 
where the quantity is defined by $
\k^{\cal {I \, J \, K \, M}} \equiv \delta^{IM}\delta^{JK} - \delta^{IK}\delta^{JM}
$.  We should mention that these were obtained by applying the 0-brane
reduction procedure to the the results presented in \cite{N4YM1}. 

The super-commutator for the singlet-supercovariant ${\rm D}_a$ with the 
triplet-supercovariant ${\rm D}_a^{\cal I}$ takes the form
\be\label{eq:crosstermsspecific}\eqalign{
\{{\rm D}_a, {\rm D}_b^{\cal I}\} A^{\cal J} ~=~ & i \, 2\,  \epsilon^{\cal {I 
\, J \, K}}\, C_{ab} \, \pa_{\tau} F^{\cal K}  \cr
\{{\rm D}_a, {\rm D}_b^{\cal I}\} B^{\cal J} ~=~ &i\, 2\, \epsilon^{\cal {I \,
 J \, K}}\, C_{ab} \, \pa_{\tau} G^{\cal K} \cr
\{{\rm D}_a, {\rm D}_b^{\cal I}\} F^{\cal J} ~=~ &i\, 2\,  \epsilon^{\cal {I \, J \, K}} \, C_{ab} \,\partial_{\tau} A^{\cal K} \cr
\{{\rm D}_a, {\rm D}_b^{\cal I}\} G^{\cal J} ~=~ &i\, 2\,  \epsilon^{\cal {I \, J \, K}}\, C_{ab} \,\partial_{\tau} B^{\cal K} \cr
\{{\rm D}_a, {\rm D}_b^{\cal I}\} \psi_c^{\cal J} ~=~ & i \, 2\,  \epsilon^{\cal {I \, J \, K}}\, C_{ab} 
\, (\g \cdot {\cal T})_c{}^d \partial_{\tau} \psi_d^{\cal K}  \cr
\{{\rm D}_a, {\rm D}_b^{\cal I}\} {\rm d} ~=~ &0 \cr
\{{\rm D}_a, {\rm D}_b^{\cal I}\}{\vec A} ~=~ & 0  \cr
\{{\rm D}_a, {\rm D}_b^{\cal I}\}\lambda_c ~=~ &0   ~~~.
}\ee
The first five of these equations inform us that there is a triplet central 
charge ${\cal Z}^{\cal I}$ that appears algebraically as
\be
\{{\rm D}_a, {\rm D}_b^{\cal I}\}  ~=~ i \, 2\,   C_{ab} \, {\cal Z}^{\cal I}
\ee
and 
that acts on the fields of the chiral multiplet according to:
\be \eqalign{
{\cal Z}^{\cal I}  \left(A^{\cal J}  \right) ~&=~  \epsilon^{\cal {I \, J \, K}}\, 
\partial_{\tau}  \,  F^{\cal K} ~~~~~,~~ 
{\cal Z}^{\cal I}  \left(B^{\cal J}  \right) ~=~  \epsilon^{\cal {I \, J \, K}}\,  
\partial_{\tau}  \, G^{\cal K} ~~~~~, \cr 
{\cal Z}^{\cal I}  \left(F^{\cal J}  \right) ~&=~  \epsilon^{\cal {I \, J \, K}}  \,\partial_{\tau} \,  A^{\cal K} 
~~,~~ {\cal Z}^{\cal I}  \left(G^{\cal J}  \right) ~=~  \epsilon^{\cal {I \, J \, K}}  \,\partial_{\tau}  \,  
B^{\cal K} ~~, \cr
&~~~ {\cal Z}^{\cal I}  \left(\psi_a^{\cal J} \right) ~=~  \epsilon^{\cal {I \, J \, K}}\,(
\g \cdot {\cal T})_c{}^d \partial_{\tau} \psi_d^{\cal K} ~~~.
} \ee
This same triplet central charge acts very differently on the fields 
of the vector supermultiplet where we see
\be
{\cal Z}^{\cal I}  \left(\lambda_a  \right) ~=~ {\cal Z}^{\cal I}  \left( {\vec A}
\right) ~=~ {\cal Z}^{\cal I}  \left( {\rm d}  \right) ~=~ 0 ~~~.
\ee

 So based on the experience of previous examinations based on 
 valise adinkras, the 1D, L-matrices and R-matrices derived in
 chapter six do {\em {not}} correspond to an off-shell valise.

\section{Conclusion, Summary \& Prospectus}

$~~~$ In the work, we have reached a milestone of establishing 
a 1D, $N$ = 16 formulation of the 4D, $\cal N$ = 4 abelian vector 
supermultiplet realized as a valise.  At the level of representation 
theory, the formulation is expected to capture faithfully properties 
of the four dimensional theory.

We have explicitly seen that the distinction between 4D, $\cal N$
= 1 chiral and 4D, $\cal N$ = 1 vector supermultiplets 
as characterized by the length of cycles in related
L-matrices and R-matrices is retained even though only one of four
SUSY charges is realized in an off-shell manner.

This work establishes a new platform from which to explore a very
old problem, ``Do there exist a set of fields that contain those of
the on-shell 4D, $\cal N$ = 4 abelian vector supermultiplet as a
subset and allow for the off-shell realization of four spacetime
supercharges?''

There is a widely held view that the answer to this question is negative.
One of the most cited reason given for supporting this viewpoint is a
`no-go theorem' \cite{SRThm}.  We do not disagree with this result.
However, we have long asserted that one of the assumptions at
its foundation is {\em a} {\em {priori}} about dynamics.  In particular, 
the authors observe: \vskip.01in 
\indent
$~~~$ Since all spinor auxiliary fields come in pairs (one
  as the Lagrange \newline
\indent
$~~~$ multiplier of the other), the total
  Fermi dimensionality of the off-\newline
\indent
$~~~$ shell representation
  is thus determined modulo 2d (=8N) by the \newline
\indent
$~~~$  total
dimensionality of the physical Fermi fields. \vskip.01in  \noindent
If this assumption is dropped, would the result of the no-go theorem
change?  This possibility is held out even in this work itself.  It is 
in this domain of relaxing this assumption we wish to probe using 
adinkra-enabled methodology.  

We have arrived at a rather precise reformulation of the off-shell problem
strictly in terms of linear algebra.  
The problem is to find the smallest integer $p$ for which sixteen distinct 
(16 + 4$p$) $\times$ (16 + 4$p$) L-matrices can be constructed that 
satisfy two conditions: \vskip.1in \indent
(a.) they must contain the 16 $\times$ 16  L-matrix sub-blocks of
(\ref{vN4Lsingl}) - (\ref{vN4Rtrip3}), and\newline \indent
(b.) realize the ``Garden Algebra'' conditions'' in (\ref{CM9}) - (\ref{GarDNAlg2}).
\vskip.1in \noindent
We strongly suspect the result for the no-go theorem will change, at 
least at the level of valise adinkras, if the assumption mentioned
above is dropped.   Superspace methods imply that there is
necessarily a solution for $p$ = 131,068, but the challenge is to find 
smaller solutions.

Firstly, as seen in
our discussion of the 4D, $\cal N$ = 1 chiral and vector supermultiplets,
the problem of auxiliary fields is equivalent to adding more nodes
to the adinkra corresponding to the on-shell theory.  In the example
of the 4D, $\cal N$ = 1 supermultiplets, the addition of such nodes 
corresponds to enlarging the L-matrices and R-matrices (analogous 
to starting with those of \ref{chiD0N}, \ref{chiD0O}, \ref{V1D0H},
\ref{V1D0HO}) in such a way as to satisfy the  conditions in (\ref{CM9}), 
(\ref{GarDNAlg1}), and (\ref{GarDNAlg2}).

This is a problem that has similarities to a problem in cryptography 
and is familiar to anyone who has seen the television game show,
``Wheel of Fortune'' or in some versions of a crossword puzzle. 
Some letters, but not all, in words are given, and the point is reconstruct
the complete words.  The ``Garden Algebras'' in this context serves
as the dictionary of acceptable words.

In our 1D, $N$ = 4 examples, the matrices of \ref{chiD0N}, \ref{chiD0O}, 
\ref{V1D0H}, and  \ref{V1D0HO} act as the initial letters and the complete 
matrices (given in Appendix B) satisfying the conditions 
in (\ref{GarDNAlg1}) - (\ref{GarDNAlg2}) play the role of the completed 
words.  Apparently for the 1D, $N$ = 16 supermultiplet, the matrices 
of chapter six play the role of the initial letters, and the quest should be 
to complete them so that their augmentation into even larger square arrays 
satisfy the conditions in (\ref{GarDNAlg1}) - (\ref{GarDNAlg2}). 

Even if there is a solution for the valise adinkra, there would
remain the problem of lifting of the nodes and the restoration
of 4D Lorentz invariance.  So success at the level of 
adinkras is no guarantee for the full field theory.  However,
even in this case, the nature of any obstruction would be
clarified.

In future works, we will report on our continuing efforts.

 \vspace{.05in}
 \begin{center}
 \parbox{4in}{{\it ``The world will never be the same once you've seen $~~$ $~~$
 it through the eyes of Forrest Gump.''}\,\,-\,\, Forrest Gump $~~~~~~~~~$}
 \parbox{4in}{{\it ``Everything should be made as simple as possible, but not 
$~~$  simpler.''}\,\,-\,\, Attributed to A.\  Einstein $~~~~~~~~~$
 \newline
 $~~$}  
 \end{center}
 
 \noindent
{\bf Added Note In Proof}\\[.1in] \indent
 During the course of this work, it became apparent that the color
 assignments to two of the adinkra graphs in the work of \cite{G-1} are
 inconsistent with the rest of that work.  To obtain a consistent
 assignment, whenever the vector supermultiplet adinkras are
 illustrated there, the follow color reassignments need to be made:
\newline \indent (a.) orange links $\to$ purple links, 
\newline \indent (b.) green links $\to$ orange links, and
\newline \indent (c.) purple links $\to$ green links. \newline \noindent 
In our current work this inconsistency has been eliminated.
 \newpage
 \noindent
{\bf Acknowledgements}\\[.1in] \indent
We would like to acknowledge Professors Kevin Iga and 
Kory Stiffler for helpful conversations.
This work was partially supported by the National Science Foundation 
grants PHY-0652983 and PHY-0354401. This research was also 
supported in part the University of Maryland Center for String \& Particle Theory. 
Additional acknowledgment is given by M.\ Calkins, D.\ E.\ A.\ Gates, 
and B.\ McPeak to the Center  for String and Particle Theory, as 
well as recognition for their participation in 2013 SSTPRS (Student 
Summer Theoretical Physics Research Session).  Adinkras were 
drawn with the aid of the {\em  Adinkramat\/}~\copyright\,2008 by 
G.~Landweber.     \vskip.15in

\noindent
{\bf {\Large {Appendix A: $\bm {4D}$ Supersymmetry Results}}}

In our conventions, in the set of equations describing the chiral
supermultiplet in four dimensions take the forms,
$$ \eqalign{
{\rm D}_a A ~&=~ \psi_a  ~~~~~~~~~~~~\,~~, ~~~
{\rm D}_a B ~=~ i \, (\gamma^5){}_a{}^b \, \psi_b  ~~~~~~\,~~~, \cr
{\rm D}_a \psi_b ~&=~ i\, (\gamma^\mu){}_{a \,b}\, \left(  \partial_\mu A 
\right) ~-~  (\gamma^5\gamma^\mu){}_{a \,b} \, \left( \partial_\mu B \right)
~-~ i \, C_{a\, b} \,F  ~+~  (\gamma^5){}_{ a \, b} G  ~~, \cr
{\rm D}_a F ~&=~  (\gamma^\mu){}_a{}^b \, \left( \partial_\mu \, \psi_b 
\right)  ~~~, ~~~
{\rm D}_a G ~=~ i \,(\gamma^5\gamma^\mu){}_a{}^b \, \left( \partial_\mu \,  
\psi_b \right) ~~~,
} \eqno(A.1)
$$
which imply the supersymmetry property
$$
\{ ~ {\rm D}_a  \,,\,  {\rm D}_b ~\} 
~=~  i\, 2 \, (\gamma^\mu){}_{a \,b}\,  \partial_\mu  ~~~.
\eqno(A.2)
$$
Finally, the action given by
$$
\eqalign{
 \mathcal{L}_{CM} = &  -\frac{1}{2} \partial_{\mu}A \partial^\mu A -\frac{1}{2} 
 \partial_{\mu}B \partial^\mu B  + i \frac{1}{2}  (\gamma^\mu)^{bc} {\psi}_b 
 \partial_\mu {\psi}_c + \frac{1}{2}  F^ 2 + \frac{1}{2} G^2 ~~~,
} \eqno(A.3)
$$
possesses a symmetry (up to a surface term) under the variations implied by 
(A.1).  

The analogous the set of equations describing the vector
supermultiplet in four dimensions take the forms,
$$ \eqalign{
{\rm D}_a \, A{}_{\mu} ~&=~  (\gamma_\mu){}_a {}^b \,  \l_b  ~~~, \cr
{\rm D}_a \l_b ~&=~   - \,i \, \fracm 14 ( [\, \gamma^{\mu}\, , \, \gamma^{\nu} 
\,]){}_a{}_b \, (\,  \partial_\mu  \, A{}_{\nu}    ~-~  \partial_\nu \, A{}_{\mu}  \, )
~+~  (\gamma^5){}_{a \,b} \,    {\rm d} ~~,  \cr
{\rm D}_a \, {\rm d} ~&=~  i \, (\gamma^5\gamma^\mu){}_a {}^b \, 
\, \left( \partial_\mu \l_b  \right) ~~~. \cr
} \eqno(A.4)
$$
Up to a gauge transformation on the spin-1 field these also satisfy
the algebra described by (\ref{SUSYalg}).  The action given by
$$
\eqalign{
 \mathcal{L}_{VM} = &  - \frac{1}{4} F_{\mu\nu} F^{\mu\nu} 
 + i \frac{1}{2}  (\gamma^\mu)^{bc} \l_b \partial_\mu \l_c 
 + \frac{1}{2} {\rm d}^2    ~~~,
} \eqno(A.5)
$$
possesses a symmetry (up to a surface term) under the variations implied by 
(A.4).

A common treatment of the 4D, $\cal N$ = 4 vector supermultiplet \cite{N4YMo3,N4SUSYSF1,N4SUSYSF2} 
is one where the supersymmetry 
derivatives are {\em {not}} treated symmetrically.  In this asymmetrical treatment, 
one of the supersymmetric covariant derivatives (that can be denoted by D${}_a
$) is realized in an off-shell manner, while the remaining three (denoted by 
D${}_a^{\cal I}$ with $\cal I$ = 1, 2, and 3) are not treated in an off-shell manner.  

At the level of component fields, the action for a U(1) 4D, $\cal N$ = 4, 
supersymmetric vector supermultiplet includes six spin-0 bosons ($A^{\cal I}$ 
and $B^{\cal I}$), one spin-1 gauge boson ($A_{\mu}$), four spin-1/2 fermions 
($\lambda_{a}$ and $\psi_{a}^{\cal I}$), and seven auxiliary spin-0 fields 
(d, $F^{\cal I}$, and $G^{\cal I}$) and takes the form
$$ \eqalign{
{\mathcal{L}}~=~ &-\frac{1}{2}(\partial_{\mu}A^{\cal I})(\partial^{\mu}A^{\cal I}) 
-\frac{1}{2}(\partial_{\mu}B^{\cal I})(\partial^{\mu}B^{\cal I})\cr
&+i\frac{1}{2}(\gamma^{\mu})^{ab}\psi_{a}^{\cal I}\partial_{\mu}\psi_{b
}^{\cal I} +\frac{1}{2}(F^{\cal I})^{2}+\frac{1}{2}(G^{\cal I})^{2}\cr
&-\frac{1}{4}F_{\mu\nu}F^{\mu\nu} + i \frac{1}{2} (\gamma^{\mu})^{
cd}\lambda_{c}\partial_{\mu}\lambda_{d}+\frac{1}{2}{\rm d}^2  ~~~.
}  \eqno(A.6) $$
where the gamma matrices throughout our discussion are defined as in 
Appendix A of~\cite{G-1}.    This Lagrangian is invariant up to surface 
terms with respect to the global supersymmetric transformations define 
in (A.1) which are here modified to take into account there are now 
three independent chiral supermultiplets.  So we 
have
$$ \eqalign{
{\rm D}_a A^{\cal I} ~=~&~ \psi_a^{\cal I}  ~~~, \cr
{\rm D}_a B^{\cal I} ~=~ &~i \, (\gamma^5){}_a{}^b \, \psi_b^{\cal I}  ~~~, \cr
{\rm D}_a \psi_b^{\cal I} ~=~ &~i\, (\gamma^\mu){}_{a \,b}\, \, \left( \pa_{\m} A^{\cal I} 
\right) ~-~ ~ (\gamma^5\gamma^\mu){}_{a \,b} \,\, \left( \pa_{\m} B^{\cal I} 
\right) ~\cr
&-~ ~i \, C_{a\, b} 
\,F^{\cal I}  ~+~ ~ (\gamma^5){}_{ a \, b} G^{\cal I}  ~~, \cr
{\rm D}_a F^{\cal I} ~=~&~ (\gamma^\mu){}_a{}^b \,\, \left( \pa_{\m} \, 
\psi_b^{\cal I}   \right)
~~~, \cr
{\rm D}_a G^{\cal I} ~=~ &~i \,(\gamma^5\gamma^\mu){}_a{}^b \,\, 
\left( \pa_{\m} \,  \psi_b^{\cal I}  \right) ~~~,
} \eqno(A.7)
$$
under the singlet D-operator acting on the three 4D, $\cal N$ = 1 chiral 
supermultiplets.   For the 4D, $\cal N$ = 1 vector supermultiplet, the 
realization of the action of the singlet D-operator is given by (A.4) still.
In order to realize 4D, $\cal N$ = 4, a triplet D${}_a{}^{\cal I}$-operators 
is required.

There is a well-known realization of the triplet D${}_a{}^{\cal I}$-operators 
$$ \eqalign{
{\rm D}_a^{\cal I} \, A{}_{\mu} ~=~ & -(\gamma_\mu){}_a {}^b \,  
\psi_b^{\cal I}  ~~~, \cr
{\rm D}_a^{\cal I} \l_b ~=~ & ~i\, (\gamma^\mu){}_{a \,b}\,  
\left( \partial_\mu A^{\cal I}  \right) ~-~ ~ (\g^5\g^\mu){}_{a \,b} \, 
\left(\partial_\mu B^{\cal I}  \right) ~\cr
&-~i \, C_{a\, b} 
\,F^{\cal I}  ~-~ (\gamma^5){}_{ a \, b} G^{\cal I}  ~~, \cr
{\rm D}_a^{\cal I} \, {\rm d} ~=~& i \, (\gamma^5\g^\mu){}_a {}^b \, 
\, \, \left( \pa_{\m} \psi_b^{\cal I} \right) ~~~. \cr
} \eqno(A.8)
$$
$$ \eqalign{
{\rm D}_a^{\cal I} A^{\cal J} ~=~& \delta^{\cal {I \, J}}~ \l_a - 
\epsilon^{\cal {I \, J}}_{~~~{\cal K}}\,  \psi_a^{\cal K}
~~~, \cr
{\rm D}_a^{\cal I} B^{\cal J} ~=~& ~i \, (\gamma^5){}_a{}^b \, 
\left[~\delta^{\cal {I \, J}} ~ \l_b + \epsilon^{\cal {I \, J}}_{~~~{\cal 
K}}  \, \psi_b^{\cal K}~ \right]  ~~~, \cr
{\rm D}_a^{\cal I} \psi_b^{\cal J} ~=~& \delta^{\cal {I \, J}} {\big 
[}~ i \, \fracm 14 ( [\, \gamma^{\mu} \, , \,  \gamma^{\nu} \,]){}_a
{}_b \, (\,  \partial_\mu  \, A{}_{\nu} ~-~  \partial_\nu \, A{}_{\mu}  
\, ) ~+~  (\gamma^5){}_{a \,b} \,    {\rm d}~ {\big ]}  \cr
&+~\epsilon^{\cal {I \, J}}_{~~~{\cal K}} {\big [}~ i\, (\g^\mu){}_{a 
\,b}\, \, \left( \pa_{\m} A^{\cal K} \right) ~+~ (\gamma^5\gamma^\mu)
{}_{a \,b} \,\, \left( \pa_{\m} B^{\cal K} \right) ~\cr
&~~~~~~~~~~~-~ i \, C_{a\, b} \,F^{\cal K}  ~-~ (\gamma^5)
{}_{ a \, b} G^{\cal K}~ {\big ]}  ~~, \cr
{\rm D}_a^{\cal I} F^{\cal J} ~=~& ~ (\gamma^\mu){}_a{}^b \,
\partial_\mu \, {\big [} ~ \delta^{\cal {I \, J}} ~ \l_b  \, - 
\epsilon^{\cal {I \, J}}_{~~~{\cal K}} \,  \psi_b^{\cal K}~
{\big ]}   ~~~, \cr
{\rm D}_a^{\cal I} G^{\cal J} ~=~& ~i \,(\gamma^5\g^\mu)
{}_a{}^b \, \partial_\mu \, {\big  [}~-\delta^{\cal {I \,  J}}~ \l_b 
+ \epsilon^{\cal {I \, J}}_{~~~{\cal K}} \, \psi_b^{\cal K}~ {\big ]}  ~~~.
}  \eqno(A.8)
$$
that also leaves the action (A.6) invariant up to surface terms.  This 
applies to the component fields of the chiral supermultiplets and for
the  component fields of the vector supermultiplet we utilize.

\noindent
{\bf {\Large {Appendix B: $\bm {4D, \, \cal N}$ = 1 L-Matrices \& R-Matrices}}}

The work in \cite{G-1} derived the following explicit expressions for the
L-matrices and R-matrices of the chiral supermultiplet in four dimensions.
$$
\left( {\rm L}{}_{1}\right) {}_{i \, {\hat k}}   ~=~
\left[\begin{array}{cccc}
~1 & ~~0 &  ~~0  &  ~~0 \\
~0 & ~~0 &  ~~0  &  ~-\, 1 \\
~0 & ~~1 &  ~~0  &  ~~0 \\
~0 & ~~0 &  ~-\, 1  &  ~~0 \\
\end{array}\right] ~~~,~~~
\left( {\rm L}{}_{2}\right) {}_{i \, {\hat k}}   ~=~
\left[\begin{array}{cccc}
~0 & ~~1 &  ~~0  &  ~ \, \, 0 \\
~0 & ~~ 0 &  ~~1  &  ~~0 \\
-\, 1 & ~~ 0 &  ~~0  &  ~~0 \\
~ 0 & ~~~0 &  ~~0  &   -\, 1 \\
\end{array}\right]  ~~~,
$$
$$
\left( {\rm L}{}_{3}\right) {}_{i \, {\hat k}}   ~=~
\left[\begin{array}{cccc}
~0 & ~~0 &  ~~1  &  ~~0 \\
~0 & ~- \, 1 &  ~~0  &  ~~0 \\
~0 & ~~0 &  ~~0  &  -\, 1 \\
~1 & ~~0 &  ~~0  &  ~~0 \\
\end{array}\right] ~~~,~~~
\left( {\rm L}{}_{4}\right) {}_{i \, {\hat k}}   ~=~
\left[\begin{array}{cccc}
~0 & ~~0 &  ~~0  &  ~ \, \, 1 \\
~1 & ~~ 0 &  ~~0  &  ~~0 \\
~0 & ~~ 0 &  ~~1  &  ~~0 \\
~ 0 & ~~~1 &  ~~0  &   ~~0  \\
\end{array}\right]  ~~.
\eqno(A.1)
$$
and
$$
\left( {\rm R}{}_{1}\right) {}_{i \, {\hat k}}   ~=~
\left[\begin{array}{cccc}
~1 & ~~0 &  ~~0  &  ~~0 \\
~0 & ~~0 &  ~~1  &  ~~ 0 \\
~0 & ~~0 &  ~ 0  &  ~- 1 \\
~0 & -1 &  ~ 0  &  ~~0 \\
\end{array}\right] ~~~,~~~
\left( {\rm R}{}_{2}\right) {}_{i \, {\hat k}}   ~=~
\left[\begin{array}{cccc}
0 & ~~0 &  ~-\, 1  &  ~ \, \, 0 \\
~1 & ~~ 0 &  ~~0  &  ~~0 \\
~0 & ~~ 1 &  ~~0  &  ~~0 \\
~ 0 & ~~~0 &  ~~0  &   -\, 1 \\
\end{array}\right]  ~~~,
$$
$$
\left( {\rm R}{}_{3}\right) {}_{i \, {\hat k}}   ~=~
\left[\begin{array}{cccc}
~0 & ~~0 &  ~~0  &  ~1 \\
~0 & - \, 1 &  ~~0  &  ~0 \\
~1 & ~~0 &  ~~0  &  ~0 \\
~0 & ~~0 &   - 1  &  ~0 \\
\end{array}\right] ~~~,~~~
\left( {\rm R}{}_{4}\right) {}_{i \, {\hat k}}   ~=~
\left[\begin{array}{cccc}
~0 & ~~1 &  ~~0  &  ~ \, \, 0 \\
~0 & ~~ 0 &  ~~0  &  ~~1 \\
~0 & ~~ 0 &  ~~1  &  ~~0 \\
~ 1 & ~~~0 &  ~~0  &   ~~0  \\
\end{array}\right]  ~~~.
\eqno(A.2)
$$
The vector supermultiplet has the set of L-matrices and R-matrices as 
first derived in \cite{G-1} to be
$$
\left( {\rm L}{}_{1}\right) {}_{i \, {\hat k}}   ~=~
\left[\begin{array}{cccc}
~0 & ~1 &  ~ 0  &  ~ 0 \\
~0 & ~0 &  ~0  &  -\,1 \\
~1 & ~0 &  ~ 0  &  ~0 \\
~0 & ~0 &  -\, 1  &  ~0 \\
\end{array}\right] ~~~,~~~
\left( {\rm L}{}_{2}\right) {}_{i \, {\hat k}}   ~=~
\left[\begin{array}{cccc}
~1 & ~ 0 &  ~0  &  ~ 0 \\
~0 & ~ 0 &  ~1  &  ~ 0 \\
 ~0 & - \, 1 &  ~0  &   ~ 0 \\
~0 & ~0 &  ~0  &  -\, 1 \\
\end{array}\right]  ~~~,
$$
$$ {~~~~}
\left( {\rm L}{}_{3}\right) {}_{i \, {\hat k}}   ~=~
\left[\begin{array}{cccc}
~0 & ~0 &  ~ 0  &  ~ 1 \\
~0 & ~1 &  ~0  &   ~0 \\
~0 & ~0 &  ~ 1  &  ~0 \\
~1 & ~0 &  ~0  &  ~0 \\
\end{array}\right] ~~~~~~,~~~
\left( {\rm L}{}_{4}\right) {}_{i \, {\hat k}}   ~=~
\left[\begin{array}{cccc}
~0 & ~0 &  ~1  &  ~ 0 \\
-\,1 & ~ 0 &  ~0  &  ~ 0 \\
 ~0 & ~0 &  ~0  &   - \, 1 \\
~0 & ~1 &  ~0  &  ~  0 \\
\end{array}\right]  ~~~,
\eqno(A.3)
$$
and
$$
\left( {\rm R}{}_{1}\right) {}_{i \, {\hat k}}   ~=~
\left[\begin{array}{cccc}
~0 & ~~0 &  ~1  &  ~0 \\
~1 & ~~0 &  ~0  &  ~ 0 \\
~0 & ~~0 &  ~ 0  & ~- 1 \\
~0 & -1 &  ~ 0  &  ~0 \\
\end{array}\right] ~~~,~~~
\left( {\rm R}{}_{2}\right) {}_{i \, {\hat k}}   ~=~
\left[\begin{array}{cccc}
1 & ~0 &  ~0  &  ~ \, \, 0 \\
~0 & ~ 0 &  -\,1  &  ~~0 \\
~0 & ~ 1 &  ~~0  &  ~~0 \\
~ 0 & ~0 &  ~~0  &   -\, 1 \\
\end{array}\right]  ~~~,
$$
$$ {~~~}
\left( {\rm R}{}_{3}\right) {}_{i \, {\hat k}}   ~=~
\left[\begin{array}{cccc}
~0 & ~~0 &  ~~0  &  ~1 \\
~0 &   ~1 &  ~~0  &  ~0 \\
~0 & ~~0 &  ~~1  &  ~0 \\
~1 & ~~0 &   ~~0  &  ~0 \\
\end{array}\right] ~~~,~~~
\left( {\rm R}{}_{4}\right) {}_{i \, {\hat k}}   ~=~
\left[\begin{array}{cccc}
~0 & -1 &  ~0  &  ~ \, \, 0 \\
~0 & ~\, 0 &  ~0  &  ~~1 \\
~1 & ~\, 0 &  ~0  &  ~~0 \\
~0 & ~\, 0 &  - 1  &   ~~0  \\
\end{array}\right]  ~~~.
\eqno(A.4)
$$
$$~~$$
\noindent
{\bf {\Large {Appendix C: $\bm {4D, \, \cal N}$ = 4 L-Matrices \& R-Matrices}}}

Here we give the explicit results for the triplet $(\L_{\rm I}^{[0]})_{i \hat{k}}$ L-matrices
related to the $\cal N$ = 4 supermultiplet that are analogous to those appearing in 
(\ref{vN4Lsingl}) for the singlet L-matrices.
$$
\left( {\rm L}{}_{1}^{[1]}\right) {}_{i   {\hat k}}     ~=~  \left[\begin{array}{cccc}
0 			& 0 			& 0 			& (2)_b(243) \\
0			& 0 			& (15)_b (243)	& 0 \\
0			& (0)_b(243) 	& 0 			& 0 \\
(13)_b (1243)	& 0			& 0 			& 0
\end{array}\right] {      ,         }  {               }
$$
$$
\left( {\rm L}{}_{2}^{[1]}\right) {}_{i   {\hat k}}     ~=~  \left[\begin{array}{cccc}
0 			& 0 			& 0 			& (4)_b(123) \\
0			& 0 			& (9)_b (123)	& 0 \\
0			& (6)_b(123) 	& 0 			& 0 \\
(11)_b (23)	& 0			& 0 			& 0
\end{array}\right]  {           ,    }  {               }
$$
$$
\left( {\rm L}{}_{3}^{[1]}\right) {}_{i   {\hat k}}     ~=~  \left[\begin{array}{cccc}
0 			& 0 			& 0 			& (14)_b(134) \\
0			& 0 			& (3)_b (134)	& 0 \\
0			& (12)_b(134) 	& 0 			& 0 \\
(7)_b (14)	& 0			& 0 			& 0
\end{array}\right]  {         ,      }  {               }
$$
$$
\left( {\rm L}{}_{4}^{[1]}\right) {}_{i   {\hat k}}     ~=~  \left[\begin{array}{cccc}
0 			& 0 			& 0 			& (8)_b(142) \\
0			& 0 			& (5)_b (142)	& 0 \\
0			& (10)_b(142) 	& 0 			& 0 \\
(1)_b (1342)	& 0			& 0 			& 0
\end{array}\right]  {       ,        }  {               }
\eqno(C.1)
$$
with their associated R-matrices given by
$$
\left( {\rm R}{}_{1}^{[1]}\right) {}_{{\hat k} \, i}   ~=~
\left[\begin{array}{cccc}
0 			& 0 			& 0 			& (7)_b(1342) \\
0			& 0 			& (0)_b (234)	& 0 \\
0			& (15)_b(234) 	& 0 			& 0 \\
(8)_b (234)	& 0			& 0 			& 0
\end{array}\right]
$$
$$
\left( {\rm R}{}_{2}^{[1]}\right) {}_{{\hat k} \, i}   ~=~
\left[\begin{array}{cccc}
0 			& 0 			& 0 			& (13)_b(23) \\
0			& 0 			& (5)_b (132)	& 0 \\
0			& (10)_b(132) 	& 0 			& 0 \\
(1)_b (132)	& 0			& 0 			& 0
\end{array}\right]
$$
$$
\left( {\rm R}{}_{3}^{[1]}\right) {}_{{\hat k} \, i}   ~=~
\left[\begin{array}{cccc}
0 			& 0 			& 0 			& (14)_b(14) \\
0			& 0 			& (9)_b (143)	& 0 \\
0			& (6)_b(143) 	& 0 			& 0 \\
(11)_b (143)	& 0			& 0 			& 0
\end{array}\right]
$$
$$
\left( {\rm R}{}_{4}^{[1]}\right) {}_{{\hat k} \, i}   ~=~
\left[\begin{array}{cccc}
0 			& 0 			& 0 			& (4)_b(1243) \\
0			& 0 			& (3)_b (124)	& 0 \\
0			& (12)_b(124) 	& 0 			& 0 \\
(2)_b (124)	& 0			& 0 			& 0
\end{array}\right]
\eqno(C.2)
$$

We find for the $(\L_{\rm I}^{[2]})_{i \hat{k}}$ matrices
$$
\left( {\rm L}{}_{1}^{[2]}\right) {}_{i   {\hat k}}    ~=~  \left[\begin{array}{cccc}
0			& 0 			& (0)_b (243)	& 0 \\
0 			& 0 			& 0 			& (2)_b(243) \\
(15)_b (243)	& 0			& 0 			& 0 \\
0			& (13)_b(1234) 	& 0 			& 0 
\end{array}\right]  {      ,         }  {               }
$$
$$
\left( {\rm L}{}_{2}^{[2]}\right) {}_{i   {\hat k}}    ~=~  \left[\begin{array}{cccc}
0			& 0 			& (6)_b (123)	& 0 \\
0 			& 0 			& 0 			& (4)_b(123) \\
(9)_b (123)	& 0			& 0 			& 0 \\
0			& (11)_b(23) 	& 0 			& 0
\end{array}\right]   {         ,      }  {               }
$$
$$
\left( {\rm L}{}_{3}^{[2]}\right) {}_{i   {\hat k}}    ~=~  \left[\begin{array}{cccc}
0			& 0 			& (12)_b (134)	& 0 \\
0 			& 0 			& 0 			& (14)_b(134) \\
(3)_b (134)	& 0			& 0 			& 0 \\
0			& (7)_b(14) 	& 0 			& 0
\end{array}\right]   {         ,       }  {               }
$$
$$
\left( {\rm L}{}_{4}^{[2]}\right) {}_{i   {\hat k}}    ~=~  \left[\begin{array}{cccc}
0			& 0 			& (10)_b (142)	& 0 \\
0 			& 0 			& 0 			& (8)_b(142) \\
(5)_b (142)	& 0			& 0 			& 0 \\
0			& (1)_b(1342) 	& 0 			& 0
\end{array}\right]  {       ,         }  {               }
\eqno(C.3)
$$
with their associated R-matrices given by
$$
\left( {\rm R}{}_{1}^{[2]}\right) {}_{{\hat k} \, i}   ~=~
\left[\begin{array}{cccc}
0			& 0 			& (15)_b (234)	& 0 \\
0 			& 0 			& 0 			& (7)_b(1342) \\
(0)_b (234)	& 0			& 0 			& 0 \\
0			& (8)_b(234) 	& 0 			& 0
\end{array}\right]
$$
$$
\left( {\rm R}{}_{2}^{[2]}\right) {}_{{\hat k} \, i}   ~=~
\left[\begin{array}{cccc}
0			& 0 			& (10)_b (132)	& 0 \\
0 			& 0 			& 0 			& (13)_b(23) \\
(5)_b (132)	& 0			& 0 			& 0 \\
0			& (1)_b(132) 	& 0 			& 0
\end{array}\right]
$$
$$
\left( {\rm R}{}_{3}^{[2]}\right) {}_{{\hat k} \, i}   ~=~
\left[\begin{array}{cccc}
0			& 0 			& (6)_b (143)	& 0 \\
0 			& 0 			& 0 			& (14)_b(14) \\
(9)_b (143)	& 0			& 0 			& 0 \\
0			& (11)_b(143) 	& 0 			& 0
\end{array}\right]
$$
$$
\left( {\rm R}{}_{4}^{[2]}\right) {}_{{\hat k} \, i}   ~=~
\left[\begin{array}{cccc}
0			& 0 			& (12)_b (124)	& 0 \\
0 			& 0 			& 0 			& (4)_b(1243) \\
(3)_b (124)	& 0			& 0 			& 0 \\
0			& (2)_b(124) 	& 0 			& 0
\end{array}\right]
\eqno(C.4)
$$

We find for the $(\L^{[3]}_{\rm I})_{i \hat{k}}$ matrices
$$
\left( {\rm L}{}_{1}^{[3]}\right) {}_{i   {\hat k}}    ~=~  \left[\begin{array}{cccc}
0			& (15)_b(243) 	& 0 			& 0 \\
(0)_b (243)	& 0			& 0 			& 0 \\
0 			& 0 			& 0 			& (2)_b(243) \\
0			& 0 			& (13)_b (1243)	& 0
\end{array}\right]   {      ,         }   {               }
$$
$$
\left( {\rm L}{}_{2}^{[3]}\right) {}_{i   {\hat k}}    ~=~  \left[\begin{array}{cccc}
 0			& (9)_b(123) 	& 0 			& 0 \\
(6)_b (123)	& 0			& 0 			& 0 \\
0 			& 0 			& 0 			& (4)_b(123) \\
0			& 0 			& (11)_b (23)	& 0 
\end{array}\right]  {           ,    }  {               }
$$
$$
\left( {\rm L}{}_{3}^{[3]}\right) {}_{i   {\hat k}}    ~=~  \left[\begin{array}{cccc}
0			& (3)_b(134) 	& 0 			& 0 \\
(12)_b (134)	& 0			& 0 			& 0 \\
0 			& 0 			& 0 			& (14)_b(134) \\
0			& 0 			& (7)_b (14)	& 0
\end{array}\right]   {          ,    }  {               }
$$
$$
\left( {\rm L}{}_{4}^{[3]}\right) {}_{i   {\hat k}}    ~=~  \left[\begin{array}{cccc}
0			& (5)_b(142) 	& 0 			& 0 \\
(10)_b (142)	& 0			& 0 			& 0 \\
0 			& 0 			& 0 			& (8)_b(142) \\
0			& 0 			& (1)_b (1342)	& 0
\end{array}\right]  {       ,        }  {               }
\eqno(C.5)
$$
with their associated R-matrices given by
$$
\left( {\rm R}{}_{1}^{[3]}\right) {}_{{\hat k} \, i}   ~=~
\left[\begin{array}{cccc}
0			& (0)_b(234) 	& 0 			& 0 \\
(15)_b (234)	& 0			& 0 			& 0 \\
0 			& 0 			& 0 			& (7)_b(1342) \\
0			& 0 			& (8)_b (234)	& 0
\end{array}\right]
$$
$$
\left( {\rm R}{}_{2}^{[3]}\right) {}_{{\hat k} \, i}   ~=~
\left[\begin{array}{cccc}
0			& (5)_b(132) 	& 0 			& 0 \\
(10)_b (132)	& 0			& 0 			& 0 \\
0 			& 0 			& 0 			& (13)_b(23) \\
0			& 0 			& (1)_b (132)	& 0
\end{array}\right]
$$
$$
\left( {\rm R}{}_{3}^{[3]}\right) {}_{{\hat k} \, i}   ~=~
\left[\begin{array}{cccc}
0			& (9)_b(143) 	& 0 			& 0 \\
(6)_b (143)	& 0			& 0 			& 0 \\
0 			& 0 			& 0 			& (14)_b(14) \\
0			& 0 			& (11)_b (143)	& 0
\end{array}\right]
$$
$$
\left( {\rm R}{}_{4}^{[3]}\right) {}_{{\hat k} \, i}   ~=~
\left[\begin{array}{cccc}
0			& (3)_b(124) 	& 0 			& 0 \\
(12)_b (124)	& 0			& 0 			& 0 \\
0 			& 0 			& 0 			& (4)_b(1243) \\
0			& 0 			& (2)_b (124)	& 0
\end{array}\right]
\eqno(C.6)
$$


\begin{thebibliography}{99}

\bibitem{SG11D1}
E.\ Cremmer, B.\ Julia, and J.\ Scherk, Phys.\ Lett.\ {\bf {76B}} (1978) 409.

\bibitem{SG11D2}
L.\ Brink, and P. Howe, Phys.\ Lett.\
{\bf {91B}} (1980) 384.

\bibitem{RS1}
L.\ Randall, and R.\ Sundrum, ``Out of this world supersymmetry breaking,'' 
Nucl.\ Phys.\ {\bf {B557}} (1999) 79, arXiv [hep-th:9810155].

\bibitem{RS2}
L.\ Randall, and R.\ Sundrum, ``A Large 
mass hierarchy from a small extra dimension,'' \ Phys.\ Rev.\ Lett.\ 83 
(1999) 3370,  arXiv [hep-th:9905221].

\bibitem{RS3}
L.\ Randall, and R.\ Sundrum, ``An Alternative to 
compactification,'' Phys.\ Rev.\ Lett.\ {\bf {83}} (1999) 4690
[hep-th:9906064].

\bibitem{SG4dn8}
E.\ Cremmer, and B.\ Julia, ``The SO(8) Supergravity,'' Nucl.\ 
Phys.\ {\bf {B159}} (1979) 141.

\bibitem{GRana0}
S.\ J.\ Gates Jr., and L.\ Rana, ``On Extended Supersymmetric Quantum 
Mechanics,'' UMDEPP 93-194 (1994), unpublished.

\bibitem{GRana1}
S.\ J.\ Gates Jr., and L.\ Rana, ``Ultra-Multiplets: A New Representation of 
Rigid 2D, N = 8 Supersymmetry,'' Phys.\ Lett.\ {\bf {B342}} (1995) 132-137.

\bibitem{GRana2}
S.\ J.\ Gates Jr., and L.\ Rana,  ``A Theory of Spinning Particles for Large 
N-extended Supersymmetry (I),'' Phys.\ Lett.\ {\bf {B352}} (1995) 50, arXiv 
[hep-th:9504025].

\bibitem{GRana3}
S.\ J.\ Gates Jr., and L.\ Rana, ``A Theory of Spinning 
Particles for Large N-extended Supersymmetry (II),'' ibid.\ Phys.\ Lett.\ {\bf 
{B369}} (1996) 262, arXiv [hep-th:9510151].

\bibitem{GRana4}
S.\ J.\ Gates Jr., and L.\ Rana, ``Tuning the RADIO to the off-shell 
2-D Fayet hypermultiplet problem,'' arXiv [hep-th:9602072], unpublished.

\bibitem{ENUF1}
S.\ J.\ Gates, Jr., W.\ D.\ Linch, III, J.\ Phillips and L.\ Rana, Grav.\ Cosmol.\ {\bf 8} 
(2002) 96, arXiv [hep-th/0109109].

\bibitem{ENUF2}
S.\ J. Gates, Jr., W.\ D.\ Linch, III, J. Phillips, 
``When Superspace Is Not Enough,'' Univ. of Md Preprint \# UMDEPP-02-054, 
Caltech Preprint \# CALT-68-2387, arXiv [hep-th:0211034], unpublished.

\bibitem{N4YMo1}
L.\ Brink, J.\ H.\ Schwarz and J.\ Scherk, Nucl.\ Phys.\ {\bf {B121}} (1977) 77.

\bibitem{N4YMo2}
F.\ Gliozzi, J.\ Scherk and D.\ I.\ Olive, Nucl.\ Phys.\  {\bf { B122}} (1977)  253.

\bibitem{N4YMo3}
M.\ T.\ Grisaru, W.\ Siegel and M.\ Rocek, Nucl.\ Phys.\ {\bf {B159}} (1979) 429.

\bibitem{amplituH1}
N.\ Arkani-Hamed, and J.\ Trnka, ``The Amplituhedron,''  Caltech Preprint
CALT-68-2872, arXiv: 1312.2007 [hep-th].

\bibitem{amplituH2}
N.\ Arkani-Hamed, and J.\ Trnka, ``Into the Amplituhedron,''  
Caltech Preprint CALT-68-2873, arXiv: 1312.7878 [hep-th].

\bibitem{N4YM1}
S.\ J.\ Gates, Jr., J.\ Parker, V.\ G.\ J.\ Rodgers, L.\ Rodriguez, K.\ Stiffler, 
``A Detailed Investigation of First and Second Order Supersymmetries 
for Off-Shell N = 2 and N = 4 Supermultiplets,'' Univ. of MD Preprint
UMDEPP-11-009, arXiv:1106.5475 [hep-th], unpublished.

\bibitem{G-1}
S.J.\, Gates, Jr. \. J.\,Gonzales, B.\, MacGregor, J.\,Parker, R.\,Polo-Sherk, V.G.J.\, 
Rodgers and L.\, Wassink, ``4D, N = 1 Supersymmetry Genomics (I),'' JHEP {\bf 
0912}, 008 (2009), e-Print: arXiv:0902.3830 [hep-th].

\bibitem{permutadnk}
I.\ Chappell, II, S.\ J.\ Gates, Jr, and T.\ H\" ubsch, ``Adinkra (In)Equivalence From Coxeter 
Group Representations: A Case Study,''  Oct 2012. 23 pp. UMD-PP-012-014,
e-Print: arXiv:1210.0478 [hep-th].

\bibitem{adinkra1}
M. Faux, S. J. Gates Jr., ``Adinkras: A Graphical Technology for Supersymmetric 
Representation Theory,'' Phys. Rev. {\bf{D71}} (2005) 065002, [hep{}-th/0408004v1].

\bibitem{SRThm}
W. Siegel, and M. Ro\v cek, 
``On Off-shell Supermultiplets,'' 
Phys.\ Lett.\ {\bf {B105}} (1981) 275.

\bibitem{N4SUSYSF1}
M.\ T.\ Grisaru, M.\  Ro\v cek, and W.\ Siegel, ``Zero Three Loop beta Function in 
N = 4 Super Yang-Mills Theory,'' Phys.\ Rev.\ Lett.\ {\bf {45}} (1980) 1063.

\bibitem{N4SUSYSF2}
M.\ T.\ Grisaru, M.\  Ro\v cek, and W.\ Siegel, ``Superloops 3, Beta 0: A 
Calculation in N = 4 Yang-Mills Theory, Nucl.\ Phys.\ {\bf {B183}} (1981) 141.


\end{thebibliography}
\end{document}